# Universal Workflow Language and Software Enables Geometric Learning and FAIR Scientific Protocol Reporting

Robert W. Epps[1], Amanda A. Volk[2,3], Robert R. White[1], Robert Tirawat[1], Rosemary C. Bramante[1,4], Joseph J. Berry*[1,4,5]

[1] National Renewable Energy Laboratory, Golden, CO, 80401 USA

[2] Materials and Manufacturing Directorate, Air Force Research Laboratory, Wright Patterson Air Force Base, Dayton, OH, 45433 USA

[3] National Research Council, Washington D.C., 20001, USA

[4] Department of Physics, University of Colorado Boulder, Boulder, CO, 80309 USA

[5] Renewable and Sustainable Energy Institute, University of Colorado Boulder, Boulder, CO, 80309 USA

* Corresponding author: Joe.Berry@nrel.gov

## Abstract

The modern technological landscape has trended towards increased precision and greater digitization of information. However, the methods used to record and communicate scientific procedures have remained largely unchanged over the last century. Written text as the primary means for communicating scientific protocols poses notable limitations in human and machine information transfer. Therefore, successful replication and analysis of experimental protocols necessitates a new approach for high-fidelity communication. In this work, we present the Universal Workflow Language (UWL) and the open-source Universal Workflow Language interface (UWLi). UWL is a graph-based data architecture that can capture arbitrary scientific procedures through workflow representation of protocol steps and embedded procedure metadata. It is machine readable, discipline agnostic, and compatible with Findable, Accessible, Interoperable, and Reusable reporting standards. UWLi is an accompanying software package for building and manipulating UWL files into tabular and plain text representations in a controlled, detailed, and multilingual environment. UWL transcription of protocols from three high-impact publications resulted in the identification of substantial deficiencies in the detail of the reported procedures. UWL transcription of these publications identified seventeen procedural ambiguities and thirty missing parameters for every one hundred words in published procedures. In addition to preventing and identifying procedural omission, UWL files were found to be compatible with geometric learning techniques for representing scientific protocols. In a surrogate function designed to represent an arbitrary multi-step experimental process, graph transformer networks were able to predict outcomes in approximately 6,000 fewer experiments than equivalent linear models. Implementation of UWL and UWLi into the scientific reporting process will result in higher reproducibility between both experimentalists and machines, thus proving an avenue to more effective modeling and control of complex systems.

## Introduction

Advances in every field of empirical scientific research, from biology to materials science, are stalled by challenges in reproducibility. Even within a peer-reviewed setting, the methods with which experimental protocols are carried out, recorded, and communicated often fail to capture the precision and accuracy necessary for effective knowledge transfer of technological advances. This reproducibility challenge, well known across disciplines, impairs the rate of communal scientific discovery and wastes resources.[1–5] Consequently, scientific solutions are generated at a slower rate and higher cost than necessary. The detriments of such effects are especially pronounced in research fields targeting urgent technological needs, such as energy generation and storage.

Specific to energy generation, photovoltaic (PV) technologies often struggle with consistent reproduction of laboratory scale protocols. With the needed thin-film PV research boom has come frequent reporting of champion devices, without verified reproducibility or comprehensive protocols. Systemic study of these properties in conjunction with thin-film PV development, particularly for metal halide perovskite thin-films, may provide insight into mechanisms of improvement. Prior work has attributed the challenges in replicability with the high sensitivity and complexity of PV materials coupled with incomplete procedure reporting.[6] When experimental procedures consistently cannot be effectively transferred between two domain experts, the efficacy of peer-reviewed, literature-driven communication comes into question. A lack of control at the lab-scale makes functional application highly challenging when the target goal is effective scale up and rollout. Consequently, many technologies suffer from a significant time delay and cost barrier between laboratory demonstration and implementation in the broader world. In the context of meaningful application, robust protocols are just as important as high performing materials and devices, so there is considerable need for effective, high-fidelity representation of developed procedures.

One aspect of reproducibility issues can be attributed to imprecise and inconsistent natural language-based reporting of protocols. Shown in Figure 1A, the language and detail typically found in peer-reviewed manuscripts often fails to capture sufficient detail for immediate replication. While incomplete procedure descriptions could be potentially mitigated through fieldwide enforcement of minimum detail reporting standards, historical adoption of standards has relied on accessibility. Computer science, for example, has featured notable success in generating high reproducibility research due to the facile means with which code can be shared and reproduced.[7,8] Within the peer-review process, it is simple to request the upload of the source code used to generate a data set. Similarly, Findable, Accessible, Interoperable, and Reusable (FAIR) reporting standards have been implemented through various fields of physical sciences through data repositories.[9–11] However, these systems focus on the standardized formatting and uploading of well-defined data sets. For example, the Cambridge Crystallographic Data Center (CCDC) provides a database and guidelines for describing and distributing characterized structural data.[12] Like the other commonly employed data repositories, the CCDC operates within the scope of definitive measurements on synthesized materials. To date, there is not a universally FAIR compatible data format for recording the steps taken in non-computational experimental procedures. While there is a collection of interactive laboratory notebooks tools that enable enhanced science communications,[13] it is unlikely that uniform and comprehensive protocol reporting will be universally implemented if it relies on undirected natural language.

Beyond the limitations of natural language protocol reporting, most efforts to record and store procedural data rely on inefficient data architectures. The current paradigm for databasing scientific procedures trend towards tabulated data entry, likely due to the prolific availability of tabulating software. However, relational database formats are not conducive to capturing the inherent structure of experimental procedures, and forcing procedure data into these formats can incur information loss. For example, shown in Figures 1B and 1C, relational databases struggle to correlate the entered fields in a meaningful and complete manner. The table fails to account for sequence dependent steps in the procedure, and it does not differentiate between missing entire process steps and or simply missing protocol details. Tabular formats can, in theory, capture any arbitrary protocol given enough parameters and consistency in data entry. However, to compare large and diverse multistep processes in a meaningful context, a high degree of specificity is required. When procedures regularly take well over twenty process steps, can be conducted in a wide variety of sequences, and often feature completely novel process components, the complexity of capturing every process in the same relational database quickly becomes overwhelming.

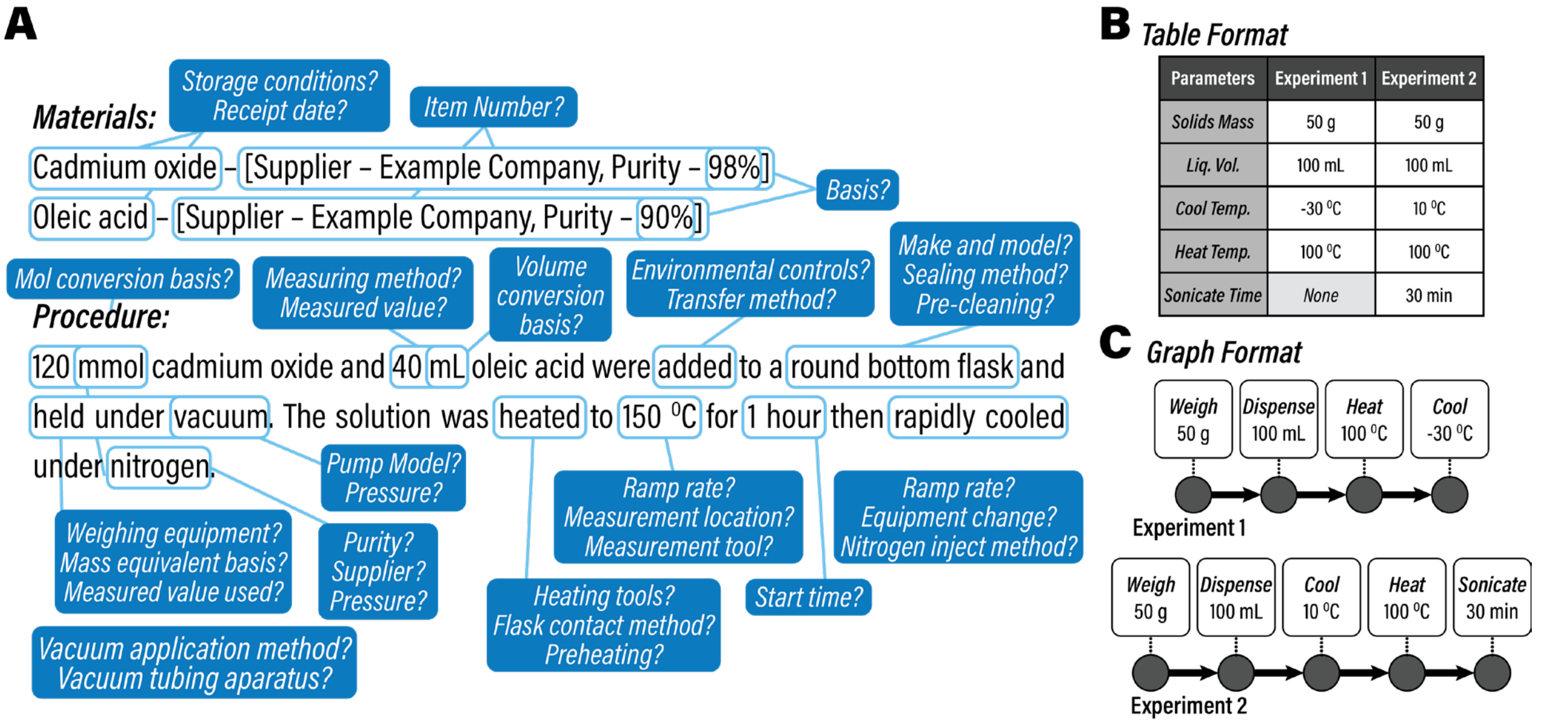

**Figure 1. Illustration of current challenges in procedure recording. (A) An example segment of an experimental procedure with annotated ambiguities. Illustrations of data formatting on a set of four arbitrary experimental procedures in (B) tabular and (C) graph formats.**

These challenges in effective data capture are evident in many efforts to systematically record human-led experimental procedures. The Perovskite Database (PDB), for example, is a manually compiled library of every accessible perovskite photovoltaic device published in peer-reviewed literature,[14] and even within this constrained context and within a field with established measurement standards,[15,16] the diversity of approaches is overwhelming.[17] The PDB features more than 400 parameters across over 40,000 logged devices. Within these entries, 75% of all fields in the complete database contain no information. However, most protocols arguably could not be accurately reproduced from their table entry alone. To address this issue, researchers continuously update the database features with relevant parameters.[14] While useful in improving the specificity, this approach does not provide a long-term solution as fields develop and new information is made available, and it requires constant revision of prior entries. Additionally, data driven insights are constrained to process features that humans have already identified as relevant, impeding opportunities for machine learning driven exploration. Despite a massive and coordinated undertaking that likely recorded almost all available information in literature, it is challenging to derive new research directions from the data set in its current format.[18–21] Extend these challenges beyond a single application of a single material and across scientific research in general, and the difficulty of the undertaking grows exponentially.

The current standard of written English procedure reporting provides limited accessibility among both machines and humans. Protocols are published using inconsistent conventions and standards, and they are distributed non-uniformly across various journals in supporting information sections, main texts, and reference chains from prior literature. As such, the physical scientific community is underprepared in data collection strategies for the introduction of machine and algorithm directed research. Natural language processing algorithms will suffer from similar limitations as human researchers when attempting to interpret written protocols, so there is a great need for a more comprehensive and machine compatible format for algorithm guided exploration. Some efforts have been made to capture experimental procedures in a universal, machine-readable structure; however, existing studies have developed systems that either are specific to a robotic architecture, fail to capture material and equipment metadata, and/or constrain experimental procedures to a set of predetermined action functions and key item roles.[22–24] For example, the Open Reaction Database captures essential process information such as reactive components and basic processing conditions, but it is unable to capture high specificity protocol steps outside of natural language notes. Similarly, NOMAD, while it can capture process details and metadata, relies on ontological definitions of domain specific components like “spin-coating” or “solvent”. The inherent nature of a field-specific ontological approach constrains the flexibility of researchers within that field and impairs cross-field knowledge transfer. As a result, protocols cannot be captured flexibly and without significant data loss across different researchers, systems, laboratories, and fields. Consequently, systems that utilize these non-universal approaches risk isolation among incompatible research clusters. Additionally, this same challenge is present in machine-driven research efforts, including modularized robotic systems, delocalized robotic networks, and autonomous data generation systems.[25–29] While interoperability exists within a single library of robots, additional development is required for robotics to communicate protocols outside of that network. There is no existing data format for capturing comprehensive procedure metadata that is universal, complete, and discipline agnostic.

These challenges exist within the scope of knowledge transfer for a single written language, as the research community has mostly agreed that English operates as the principal language of science communication.[30] This consensus has significantly unified the larger body of scientific literature and has prevented fragmentation of research communities; however, approximately 94% of the Earth's population does not speak English as a first language, and roughly 74% do not speak English at all. The current paradigm of scientific protocol reporting presents a considerable barrier of entry for non-native English speakers and thereby reduces the net volume of scientific contributions worldwide.

Within scientific research, there is an abundance of information that could provide considerable impact if communicated both within and across disciplines, but there currently is an absence of tools that are able to consistently capture procedures in a meaningful and consistent manner. The existing methods for sharing scientific protocols lack precision, generalizability, and accessibility, inhibiting the progression of human knowledge generation. In response, we present the Universal Workflow Language (UWL) and Universal Workflow Language Interface (UWLi). UWL and UWLi are a novel, graph-based data saving format and interactive software, respectively, designed to more effectively record and report scientific procedures than the current standards of written natural language and currently available data formats. The inherent approach of UWL is to deconstruct complex processes into a series of compact, universal, and unambiguous steps that are paired with high detail metadata. Recent work has demonstrated that sequential representation of actions conducted in a protocol is among the most efficient for procedure communication and algorithmic text mining.[31] Contrasting with conventional written language protocol reporting which often seeks to summarize a process and provide base minimum detail for reproduction, UWL aims to isolate the root structure of scientific procedures into items and manipulation of those items through simple actions. UWL is a FAIR data structure for reporting scientific procedures from any research discipline.

UWLi features a visual workspace for transcribing procedures into the UWL format, parameter tabulation and management, and plain text transcription of experimental procedures. Additionally, it features an infrastructure for building multi-lingual support across global written languages in a format that enables reversible translation of experimental procedures. We evaluated the UWL architecture in a simulated graph learning scenario, where graphical representations significantly outperformed all tested conventional models in accurately representing simulated experimental procedures. Application and development of UWL and UWLi among the greater scientific community could result in more thorough, standardized, and reproducible experimental protocol reporting with greater functionality in algorithm-guided understanding of complex processes.

## Results

### *Design and Software*

UWL is a workflow data saving format that is structured in JavaScript Object Notation (JSON) and can capture any arbitrary scientific experimental procedure through a directed acyclic graph-inspired data architecture. At its root structure, a scientific procedure is a collection of actions carried out by a researcher or system on a set of tools and materials. For each action and item, there is important information on aspects of the procedure such as process parameters or resource acquisition methods. UWL is structured to capture this same root structure. The data system operates through the connection of metadata-containing nodes that represent the various materials, tools, and actions involved in a procedure with varying edge types to form distinct steps in a scientific process, as shown in Figure 2A. Each node is classified as either an action or an item node, and the UWL format is simply an associative array containing these node objects with accompanying metadata. Each node contains, among other parameters, a set of arbitrary property parameters that are defined by the user. In the case of item nodes, these properties can represent features like the material or equipment supplier, item or lot numbers, and the date of receipt or last calibration. In the case of action nodes, these parameters can represent critical process features like the setpoint temperature or the mixing time.

A step in a process is defined by an action node that is connected to a set of item nodes to form a protocol statement constructed by a simple rule set. For example, in Figure 2B, the first action node in the sequence – i.e. the first circle labeled *Add* – is connected to a set of three item nodes – i.e. the three diamonds labeled *Beaker*, *Scale*, and *Molecular Sieves*. Based on the Add action node, the three items selected are connected through Type A, B, and C connections, respectively. The A-type connected item(s) are always the object that B-type items are being added to, and the B-type connected items are always added to the A-type item(s). The C-type connected items, which are optional, are the tools used to perform the adding process. Using this ruleset, the statement is then procedurally generated to be *Add molecular sieves to beaker using scale*. Note, grammatically specific identifiers for the three edge types have been deliberately avoided to promote multilingual compatibility.

The edge logic dictated by the *Add* ruleset are inherited by a subset of action terms, such as *Mix*, *Transfer*, and *Place*. In these cases, the A-type is the stationary item(s), the B-type is the mobile item(s) approaching the stationary item, and the C-type is the tool(s) used to facilitate the motion. Similar relationships have been developed for the *Remove* and *Modify* parent classes, where *Remove* represents any separation of two items and *Modify* represents the manipulation or modification of an item, such as heating or sonication. Specific rules and further discussions are presented in Supporting Information Section 1.7, and the full list of currently supported action terms is provided in the Github repository. The current selection of supported action terms has been deliberately constrained to encourage more specificity in process recording. Instead of using a summarizing term like *Wash*, users are encouraged to apply more specific instructions through a sequence of smaller actions, such as *Add* [solvent], *Swirl* [container], then *Remove* [solvent]. However, UWL is designed comprehensively to allow for generalizable process recording, so users are also able to incorporate unsupported actions into their procedures. In these cases, users select the desired parent logic of the node-edge relationships, then implement their own unique action term.

Contrary to conventional graph architectures, the nodes in UWL entries contain two-dimensional spatial information – defined by the *Position* parameter – to dictate the sequence in which nodes are evaluated. The sequence is defined so that nodes are interpreted first by the x-coordinates – from left-to-right – then by the y-coordinates – from top-to-bottom. This sequence has little impact on item nodes other than the arrangement of multiple items connected to one action node with the same edge type. However, for action nodes, the sequence dictates the order in which steps are meant to be carried out in a procedure. Shown in Figure 2B, arrangement of these block statements can form complete experimental protocols with both embedded metadata and a high level of procedural detail.

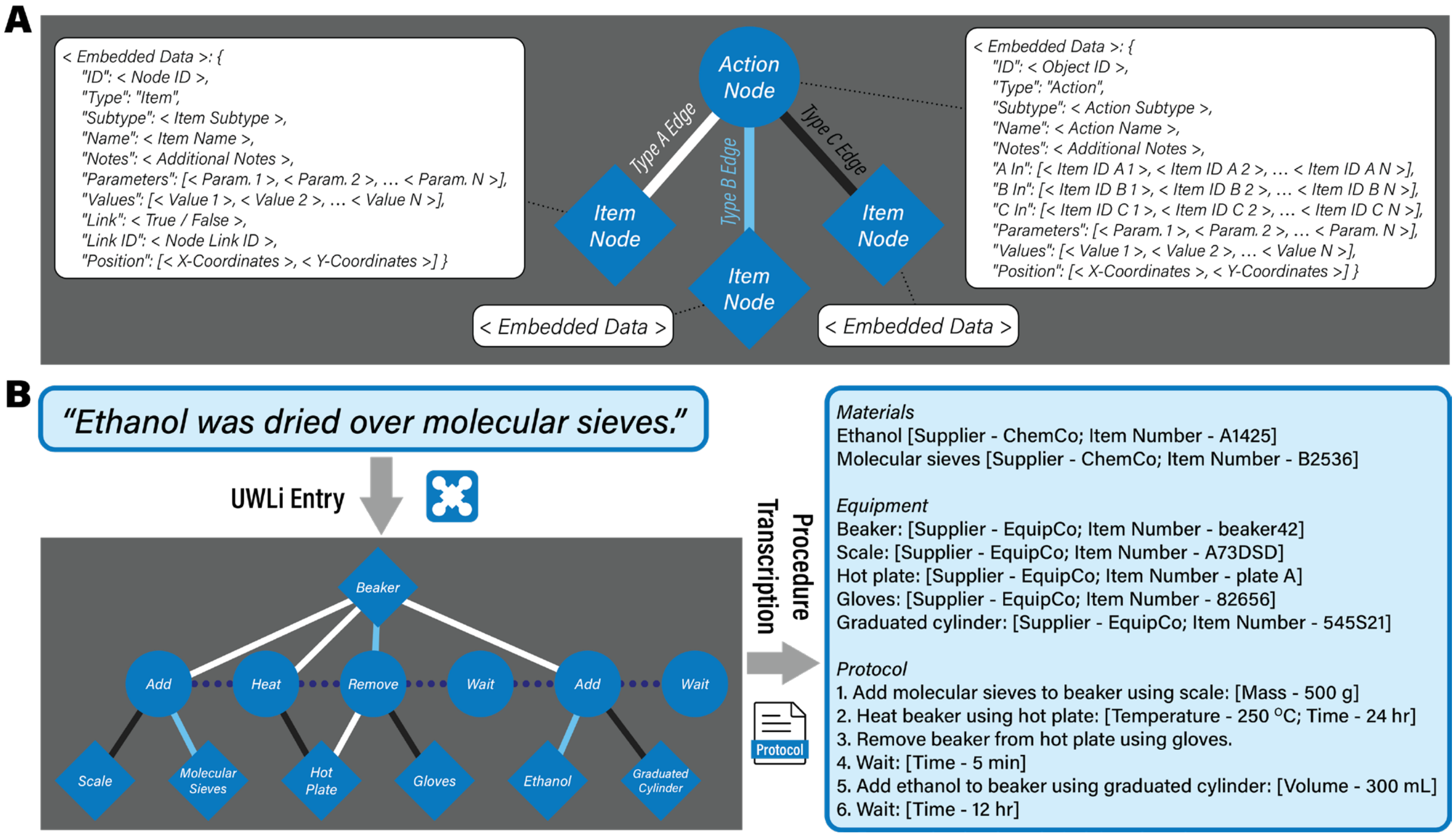

**Figure 2. Overview of Universal Workflow Language entries. (A) Illustration of metadata embedding in action and item nodes on UWL entries. (B) Example UWL entry and procedure transcription from a simple summarizing phrase, commonly used in literature reporting.**

While the ruleset specified for UWL construction is simple in design, it enables capture of highly detailed process information in a fully generalized format. Illustrated in Figure 3A, this generalizability extends into a breadth of research fields, system approaches, and data management strategies. UWL can be applied to any area of human or machine-driven research in which a specific procedure is followed, which spans from biology, physics, engineering, and computer science among many other disciplines. It also enables more effective process capture and communication for databases collected within a single research group or institution and for public-facing publications and repositories, and it provides more precise communication of protocols in human-read literature and algorithmic literature mining. Additionally, UWL enables procedure communication among different robotic experimentation systems and algorithms, machine-human integrated lab spaces, and delocalized robotic experimentation networks. UWL presents a low information loss method for universal communication of arbitrary experimental procedures.

Briefly illustrated in Figure 3B and available on GitHub, UWLi allows users to generate and modify UWL files using a drag and drop workflow building interface. The interface enables users to directly build and modify node and edge metadata, rearrange node orders, and insert external workflow segments to produce custom experimental procedures. The software features a file management system, a UWL construction workspace, automated table generation, and natural language transcription of UWL files. UWLi also provides a collection of basic usability features that are intended to streamline the protocol recording process, such as copy and paste functionality and bulk file management. Further development of these tools will improve usability, and ultimately detail, in reported procedures.

UWL files can be generated through the UWLi interface, available on GitHub, which features a full file management system, a UWL construction workspace, automated table generation, and natural language transcription of UWL files with multilingual support. Further details of the software are provided in Supporting Information Section 1. UWLi provides a framework for transcribing protocols between written languages using procedural protocol generation and translation tables. This approach, discussed in Supporting Information Section 1.10, is contrary to a model-based translation approach and enables the direct and reversable translation of process information. The current translation feature is intended as a proof of concept and is inaccurate for many spoken languages due to the algorithmic methods used to generate sentences. Human validation of the translation tables would provide more accurate transcription, and further development of these tools will improve usability, and ultimately detail, in reported procedures.

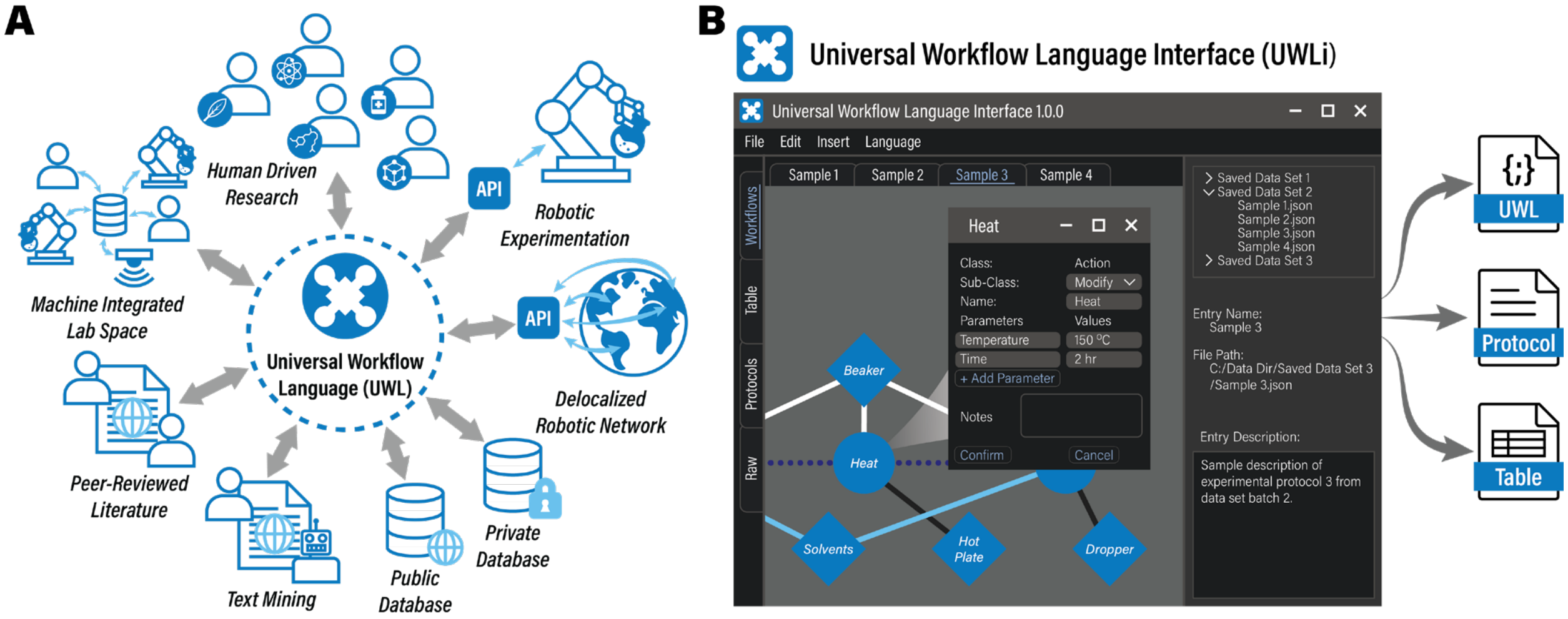

**Figure 3. Illustrations of the Universal Workflow Language and interface. (A) Illustration of the connectivity of UWL across various research fields and strategies. (B) Illustration of the UWLi used for creating, modifying, and exporting UWL files and protocols.**

### *Protocol Detail Improvement*

To compare the data density between peer-reviewed natural language and UWL, protocol excerpts from three arbitrarily selected manuscripts each recently published in the journals *Nature*, *Science*, and *Advanced Materials* were transcribed into UWL files, illustrated in Figure 4A and shown in detail in Supporting Information Section 2.[32–34] Shown in Figure 4B, by transcribing the protocols into UWL entries, significant omissions of necessary information were identified. On average, the original protocols referenced only 86% of necessary chemicals and consumable materials and 25% of necessary equipment. In addition, only 10% of the necessary parameters were given for sourcing the given materials and equipment. The most common of these parameters were suppliers and item numbers. Moreover, the original procedures contained on average only 45% of the total possible parameters identified by UWL, representing a significant data loss on consequential variables of protocols. While in some instances, the original protocols failed to report process parameters like chemical dripping rates or centrifugation speeds, in other cases, the protocols failed to fully describe the processes themselves, instead opting for summarizing action terms such as “prepared” or “washed”.

Overall, shown in Figure 4C, the UWL approach to protocol reporting with manual disparity isolation identified on average 17 procedural ambiguities for every hundred words in the original procedures. Across all tested metrics, the original protocols provided significantly fewer details on equipment, materials, and process steps. Furthermore, the original protocols, despite undergoing a rigorous peer review process, enabled considerable ambiguity, causing a high probability of irregularities during protocol reproduction.

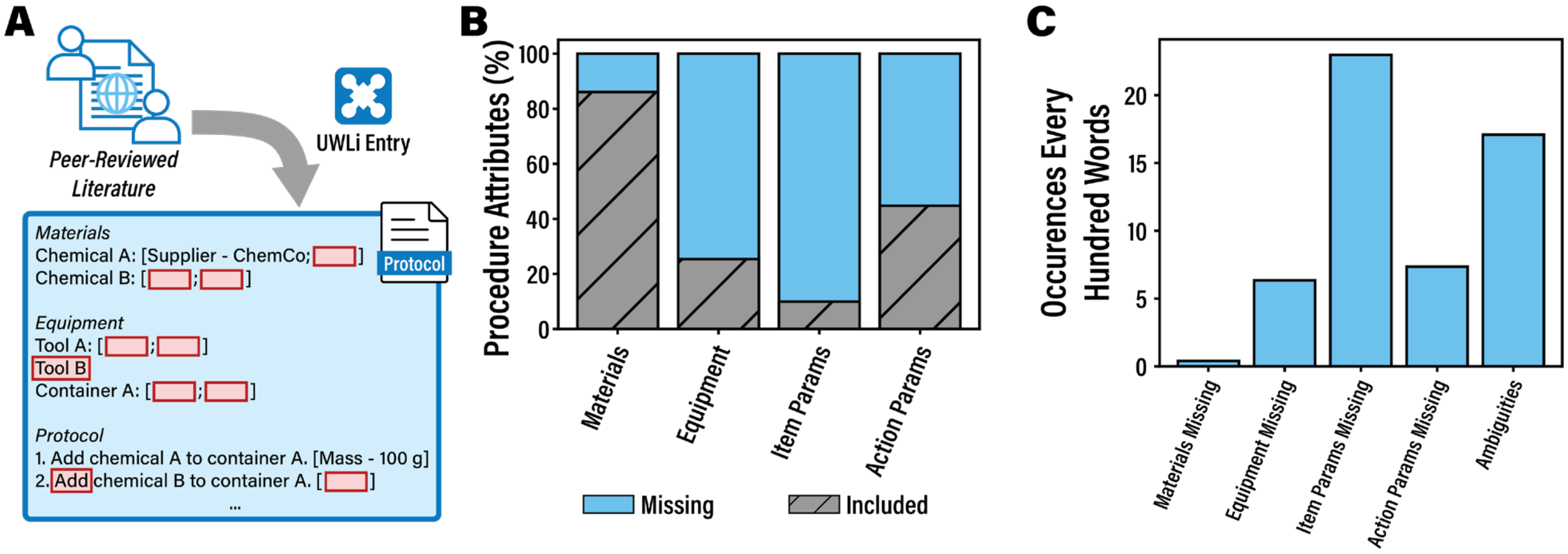

**Figure 4. Universal Workflow Language literature transcription summary. (A) Illustration of the literature transcription and flagging process. (B) The average percentage of material items, equipment items, item parameters, and action parameters included in the original procedure relative to the UWL transcribed procedure. (C) The average number of missing material items, equipment items, item parameters, and action parameters identified through UWL transcription that were not found in the original protocol and the number of manually identified ambiguities found through the UWL transcription process for every hundred words in a procedure.**

The selected protocols are by no means outliers or poor representations of protocol reporting in literature, and these procedures were identified arbitrarily from recent publications in some of the highest impact scientific journals. Lack of necessary detail is pervasive throughout scientific publications, and similar deficiencies likely can be found in almost every manuscript outside of specialty precision journals like *Organic Syntheses*.[35] For example, within the PDB 65% of the protocols provide some form of qualitative context for the active layer deposition atmosphere, and only 2.9% provide a quantitative description, despite the deposition atmosphere being a critical aspect of the film formation process.[36–38] The task of comprehensively transcribing a large and complex scientific procedure without some inbuilt structure is exceedingly difficult. The language necessary for complete communication of fine detail processes can often read as repetitive and awkward. Naturally, many scientists converge onto a protocol writing style that conveys higher language efficiency, readability and elegance, but this comes at the cost of ambiguity and incomplete reporting of details.

Simply requiring researchers to report protocols in higher detail is unlikely to completely resolve the issue. Historical precedents, such as the formation of competing unit standards and the slow introduction of FAIR reporting principles, suggest that humans are more likely to adapt systems to fit their natural inclinations than the inverse.[39] Implementation of UWL and UWLi can provide a structure for protocol reporting that enables researchers to communicate clear, complete, and precise scientific procedures more effectively. Extending this work further, UWLi provides a framework for future software that enables algorithm driven enforcement of field reporting standards.

### *Geometric Learning*

Extending beyond the features conducive to FAIR reporting principles, UWL provides a more efficient data structure for process modeling and machine learning. Recent work has shown that graph-based modeling of multi-step experimental processes can predict experimental outcomes more effectively than equivalent linear methods in many scenarios.[40] Further application of geometric learning in the experimental protocol space could lead to more efficient exploration of the complex experimental systems pertinent to the fabrication of high-performing materials and devices. UWL provides a direct connection between these new geometric modeling methods and real-world experiments. To demonstrate the viability of integrating UWL entries with geometric learning, UWL files were used to train a series of graph transformer regressor networks to predict the outcome of an arbitrary experimental surrogate function.

To generate a training data set with the surrogate function, UWL files were randomly constructed through the UWL generator, shown in Supporting Information Figure S.5. The generator creates workflows with a randomized number of actions within a specified range, with a constant number of connected items for each action. Each action node has an embedded scalar and categorical parameter, and each item node has an embedded categorical parameter. Categorical parameters represent the name of the node, and the length of the set containing all possible categorical parameters for both action and item nodes is defined by an adjustable constant. The generator randomly selects workflows under the constraints specified by the maximum number of actions, number of items per action, number of scalar parameters per action, number of possible action node names, and number of possible item node names. Through this approach, varying degrees of parameter space complexity can be evaluated.

The surrogate function, shown in more detail in the methods section of this manuscript, recursively and sequentially iterates through the items and action nodes and the associated meta data of the UWL files to generate a scalar response metric that is used for model training. The surrogate considers all embedded parameter features and the order of action steps in an interconnected design. This surrogate emulates experimental procedures in that early decisions impact the influence and outcomes of later decisions. The UWL files were converted into explicit graph representations and, coupled with the associated response values, used to train graph neural network models (GNNs), shown in Figure 5A. The GNNs use pooling of graph transformer layers followed by a linear network to generate graph-level scalar prediction of the response, which was constructed using PyTorch-Geometric[41,42] and an architecture from prior studies. [40]

The GNNs in both single and ensemble configurations were benchmarked with two typical linear models, a neural network and a gradient boosted decision tree regressor, across three levels of generator complexity – low, mid, and high. The linear model inputs were constructed through padding of categorical variables and action steps out to the maximum number of possible actions. Linear model hyperparameters were evaluated over an exhaustive grid search using models from the scikit-learn library.[43] The low, mid, and high complexity generator parameters and complexity metrics are shown in supporting information Table S.1 and S.2 respectively. The high complexity surrogate has up to seven scalar variables, forty-two categorical variables, and on the order of $10^{25}$ possible categorical combinations, making it a highly challenging and high dimension space for expensive sampling scenarios.

Shown in Figures 5B-D, the GNNs provided a substantial improvement in the test set model predictability for most training set sizes, and ensemble GNNs provided an even greater improvement to the model performance for almost all tested conditions. The gradient boosted decision tree performed similarly to the linear neural network for the low complexity space; however, the linear neural network improved at a faster rate in the mid and high complexity spaces. Relative to the ensemble GNN with a 6,400-sample training set size, the linear neural network and gradient boosted decision tree models had on average 3.2- and 5.9-times higher test set mean squared errors, respectively. For the high complexity space, the ensemble GNN achieved mean squared errors comparable to the final linear neural network and gradient boosted decision tree with approximately 5,400 and 6,200 fewer training samples, respectively. GNNs provided substantial efficiencies in modeling UWL data sets within simulated environments. Further application of UWL in experimental settings could lead to data driven insights on larger, more complex experimental spaces.

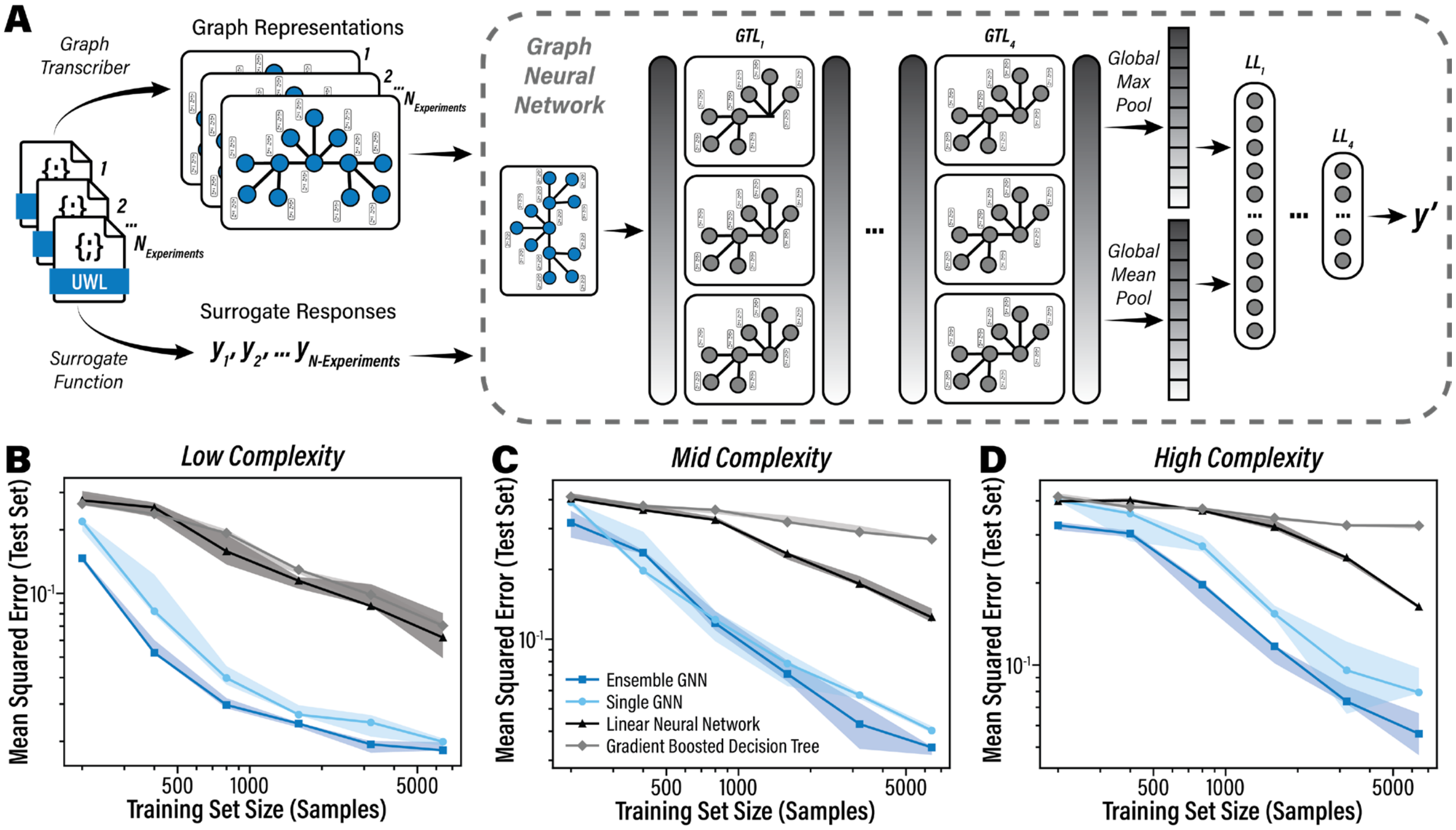

**Figure 5. Summary of surrogate modeling with graph and linear models. (A) The training architecture for generating graph data sets and training a graph neural network, comprised of graph transformer layers (GTL) and linear neural network layers (LL), for graph level prediction. The median mean squared error of a 200-sample test set as a function of training set size for an ensemble graph neural network (GNN), a single GNN, a linear neural network, and a gradient boosted decision tree across (B) low, (C) mid, and (D) high complexity workflow generators. Workflows generated with low, mid, and high complexities had maximum action lengths of 3, 5, and 7, with 1, 3, and 5 items per action, with 5, 7, and 9 total possible action types, and with 3, 5, and 7 total possible item types respectively.**

## Discussion

The data format and user interface presented in this manuscript help to alleviate several critical challenges in scientific communications. The current paradigm of manually written unstructured protocol reporting has led to substantial deficiencies in the accuracy and fidelity of scientific experimental procedures. Peer-reviewed manuscripts are consistently failing to provide sufficient detail for successful reproduction, and as a result scientific progress and ultimate application is significantly slowed and comes at a greater cost. UWL and UWLi help to support researchers in more precise and comprehensively described experimental processes, and as a result will likely facilitate more effective knowledge transfer between humans. Furthermore, UWL provides a machine compatible interface between researchers, or robotics in a lab setting, and experimental exploration and optimization algorithms. The high density of the data generated in UWL files coupled with the inherent advantages of graphical representation of experimental processes enables more efficient algorithm driven understanding of high complexity processes and more accurate databasing of complex, multi-step procedures. The development, implementation, and adoption of UWL and UWLi in all experimental research spaces provides an avenue to more effectively develop and share scientific procedures, to more accessible data for both humans and machines, and to faster advancement of science and technology.

## Experimental Procedures

### *Universal Workflow Language Structure*

UWL entries are formatted as nested associative arrays, more specifically as nested python dictionaries in the current implementation, where each node and accompanying edge characteristics and process metadata are represented by their own associative array. Parent window objects that contain the nodes then provide key value pairs that point to each node array within the workflow. A detailed description of the UWL architecture is included in Supporting Information Sections 1.1 to 1.6. Using the network of connected nodes, protocol steps are constructed using predefined action term configurations, shown in Supporting Information Figure S.2, when item nodes are connected to a single action node using combinations of the three edge types. To form an intuitive logic to the formation of protocol steps, action terms are grouped into three clusters – *add*, *remove*, and *modify* – which each interpret edge connections using a similar rule set.

The final UWL arrays are saved as JavaScript Object Notation (JSON) formatted files. With minor modifications, these structures could be transcribed to a markup language or other similar structure to incorporate additional functionalities.

### *Universal Workflow Language Interface Design*

UWLi was developed using PyQt5, a python binding set for the Qt5 application development library. Field autocompletion values were generated using ChatGPT-3.5 then curated manually, and translations were procedurally generated using Microsoft Azure Translation API.

## Data Availability

The literature review UWL files associated with this manuscript are available for download at https://github.com/NREL/Universal-Workflow-Language-Interface. A detailed description of the collected literature review data is available in the supporting information.

## Code Availability

All code associated with the Universal Workflow Language interface is available at https://github.com/NREL/Universal-Workflow-Language-Interface.

**Acknowledgements**

This work was authored in part by the National Renewable Energy Laboratory, operated by Alliance for Sustainable Energy, LLC, for the US Department of Energy (DOE) under contract no. DE-AC3608GO28308 in the Advanced Perovskite Cells and Modules program (Agreement Number 38256). Financial support for A.A.V. was provided in part by the appointment to the Air Force Science and Technology Fellowship Program at Air Force Research Laboratory, administered by the Fellowships Office of the National Academies of Sciences, Engineering, and Medicine.

**Author Contributions**

J.J.B. and R.W.E. acquired funding for the work. J.J.B and R.W.E. conceptualized the universal workflow language interface project. R.W.E. and R.R.W. designed the universal workflow language architecture. R.W.E. programed the universal workflow language interface software. R.T. and R.C.B. led universal workflow language interface user testing. A.A.V. conceptualized geometric learning on universal workflow language represented systems. A.A.V. and R.W.E. designed and programed the surrogate modeling studies. R.W.E. drafted the manuscript. All authors revised and edited the manuscript.

*Supporting Information*

# Universal Workflow Language and Software Enables Geometric Learning and FAIR Scientific Protocol Reporting

Robert W. Epps[1], Amanda A. Volk[2,3], Robert R. White[1], Robert Tirawat[1], Rosemary C. Bramante[1,4], Joseph J. Berry*[1,4,5]

[1] National Renewable Energy Laboratory, Golden, CO, 80401 USA

[2] Materials and Manufacturing Directorate, Air Force Research Laboratory, Wright Patterson Air Force Base, Dayton, OH, 45433 USA

[3] National Research Council, Washington D.C., 20001, USA

[4] Department of Physics, University of Colorado Boulder, Boulder, CO, 80309 USA

[5] Renewable and Sustainable Energy Institute, University of Colorado Boulder, Boulder, CO, 80309 USA

* Corresponding author: Joe.Berry@nrel.gov

**Table of Contents**

## 1. Universal Workflow Language Architecture

### *1.1. Sections and Root Workflows*

The highest level in a UWL entry is the root workflow. Workflows are metadata containing objects that store the individual node data as a list of nested metadata containing objects. Additionally, workflows can store nested workflows, called sections, in the object list for organizational purposes. Typically, a section corresponds to a specific phase in an experimental procedure, but sections can be allocated as needed. The root is a workflow that is not nested within any other workflow and is therefore the sole starting point for a UWL file. The metadata stored within a root workflow and section workflow follow the JSON format shown below. In addition to the object list, the root format contains basic information parameters used to classify the entire logged experiment including name, file path, and a basic text description. The section format contains similar features with the addition of parameters that position the section within its parent workflow – i.e. the object identifier and the position. The object identifier is a unique integer value used to distinguish the section from other sections and nodes throughout the entry. The position is a set of xy-coordinates used to indicate the temporal order in which objects are read in the parent workflow. Priority in this coordinate system is given to the x-value, then the y-value, meaning that objects are read from top to bottom then left to right.

### *1.2. Root Workflow Structure*

{

"Name": < Experiment Name >,

"Type": "Root",

"File": < UWL File Path ("*.json") >,

"Description": < Experiment Description >,

"Objects": {< Node $_1$ >, < Node $_2$ >, … < Node $_N$ >,

< Section $_1$ >, < Section $_2$ >, … < Section $_N$ >}

}

### *1.3. Section Structure*

< Object ID >: {

"ID": < Object ID >,

"Type": "Section",

"Name": < Section Name >,

"Description": < Section Description >,

"Objects": {< Node $_1$ >, < Node $_2$ >, … < Node $_N$ >,

< Section $_1$ >, < Section $_2$ >, … < Section $_N$ >},

"Position": [< X-Coordinates >, < Y-Coordinates >]

}

### *1.4. Action and Item Node Metadata*

The steps in the experimental protocol are comprised of metadata containing nodes and edges grouped within specific orientations. The nodes represent either items or actions affiliated with the procedure and are subdivided into respective classes, and like sections, nodes are stored as objects within a workflow object list along with basic metadata like name, description, and various identifiers. Metadata within the nodes may be modified in the interface through interactive data entry popup boxes, shown in Figure S.1. Item nodes are further classified into four sub-types – container, tool, source, and abstract. Item sub-types are primarily used for organizational purposes along with portions of the protocol transcription, discussed in a later section of the manuscript. Precise classification of items into sub-types is less critical than other steps in the recording process; however, for clarity it is most effective to classify items by their primary role in the process. Containers are used to represent any item within a procedure that contains other items, for example laboratories spaces, flasks, or well plates. Tools are items that are used to modify or interact with another item, such as hot plates, stir bars, or measurement equipment. Sources are the starting materials, typically received directly from a manufacturer. These items are generally consumed or modified as the primary subject in the process of an experiment. Abstract items are typically reserved for either embedded metadata not directly describing the process like user information or for non-physical features like etched serial numbers.

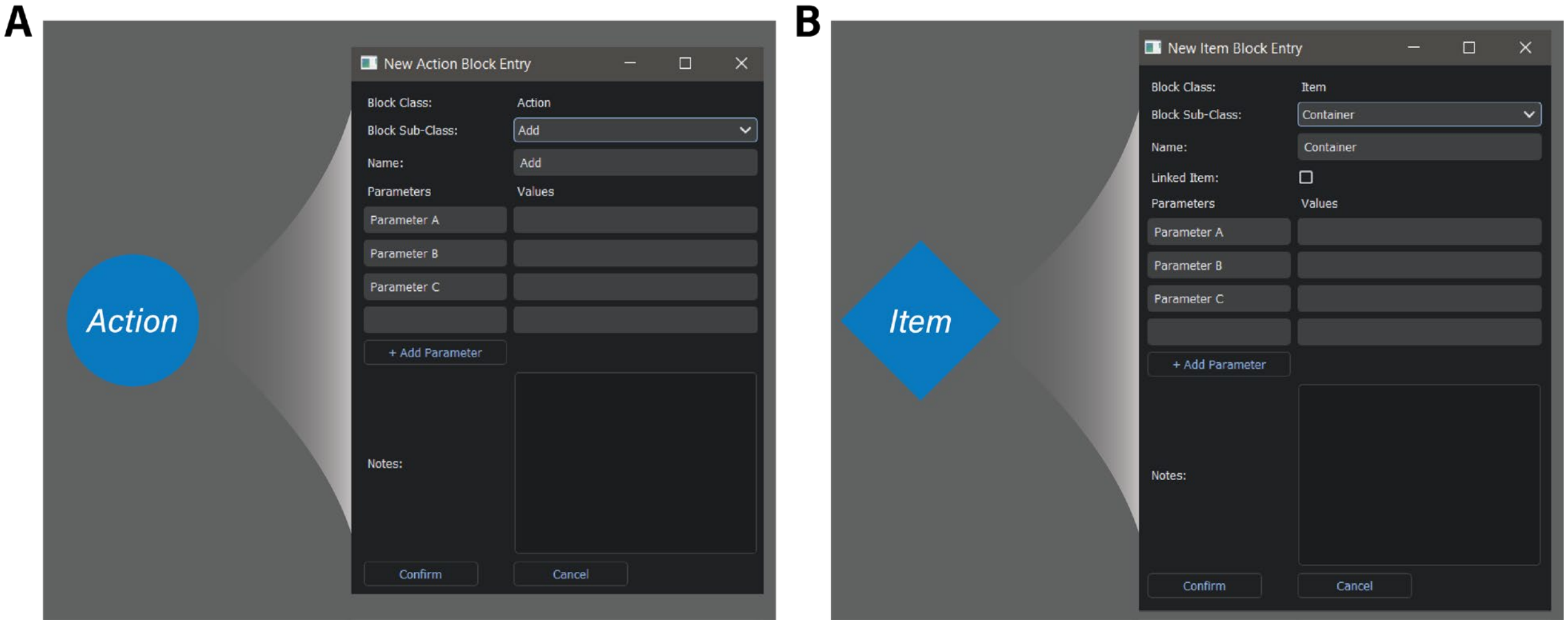

**Figure S.1. Metadata popup window within the Universal Workflow Language interface for (A) Action and (B) Item nodes.**

### *1.5. Item Node Structure*

< Object ID >: {

"ID": < Object ID >,

"Type": "Item",

"Subtype": < Item Subtype >,

"Name": < Item Name >,

"Notes": < Additional Notes >,

"Parameters": [< Parameter $_1$ >, < Parameter $_2$ >, … < Parameter $_N$ >],

"Values": [< Value $_1$ >, < Value $_2$ >, … < Value $_N$ >],

"Link": < True / False >,

"Link ID": < Object Link ID >,

"Position": [< X-Coordinates >, < Y-Coordinates >]

}

### *1.6. Action Node Structure*

< Object ID >: {

"ID": < Object ID >,

"Type": "Action",

"Subtype": < Action Subtype >,

"Name": < Action Name >,

"Notes": < Additional Notes >,

"A In": [< Item ID A $_1$ >, < Item ID A $_2$ >, … < Item ID A $_N$ >],

"B In": [< Item ID B $_1$ >, < Item ID B $_2$ >, … < Item ID B $_N$ >],

"C In": [< Item ID C $_1$ >, < Item ID C $_2$ >, … < Item ID C $_N$ >],

"Parameters": [< Parameter $_1$ >, < Parameter $_2$ >, … < Parameter $_N$ >],

"Values": [< Value $_1$ >, < Value $_2$ >, … < Value $_N$ >],

"Position": [< X-Coordinates >, < Y-Coordinates >]

}

*1.7. Edge Data and Step Construction*

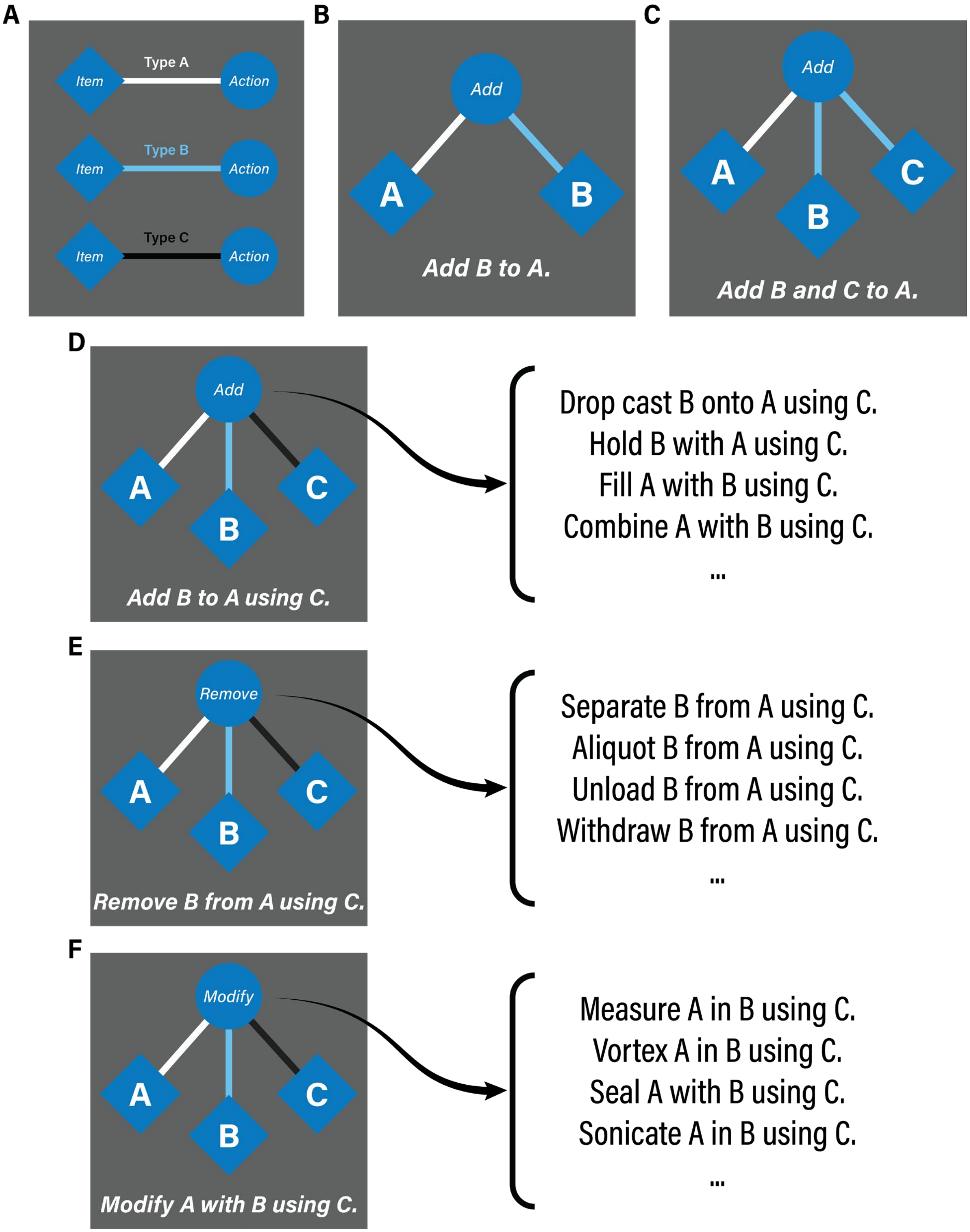

**Figure S.2. Illustration of action step construction system. (A) Summary of the three edge types, illustration of (B) a two component Add step and (C) the compound item variant, and basic illustrations of (D) Add, (E) Remove, and (F) Modify action steps with a selection of the associated child actions that inherent the parent behaviors.**

### *1.8. Organizational Tools*

The interface features a set of basic usability tools to streamline protocol transcription into UWL files. These features include basic edit, delete, copy, and paste functionality in the workflow interface. Additionally, there are standard file saving and inserting tools to enable the incorporation and modification of separate files in a single UWL. To allow for the users to use multiple item blocks across different sections or within a larger workflow, the item blocks include a link features. Blocks may be linked by assigning a common link identifier. Any two item blocks with the same link identifier will share the same metadata and indicate their linkage in the UWL data.

### *1.9. Interface Transcriptions*

Within the interface, the data represented in the Workflow tab closely represents the intended UWL architecture shown in the Raw tab and in the saved UWL .json files. To enhance usability, the Workflow tab data is transcribed programmatically into the Table and Protocol tabs. The transcription into the Protocol tab is currently not reversable Changes made in the Workflow or Table tabs are made visible in the Protocol tab, but the Protocol tab cannot be used to write or edit information in the UWL data. All blocks and metadata in a workflow are represented in the transcribed protocol, shown in Figure S.3. First, the base experimental description and name are shown at the header of the protocol. Then, the item blocks are sorted by their subclass identifier – abstract, source, material, or container – and displayed with their associated parameters and values in order of sequence in the workflow. Any notes within the item block are displayed next to the metadata listing. Finally, the action blocks and sections are iterated sequentially to build out the Procedure section. Each action block is transcribed into a single protocol step using the connected item blocks under the previously described rule set. Any parameters are notes associated with the protocol step are displayed on the same line. Each section block and its contained action blocks are assigned a subsection of the procedure, indicated by the sub-header and indentation for the nested action blocks.

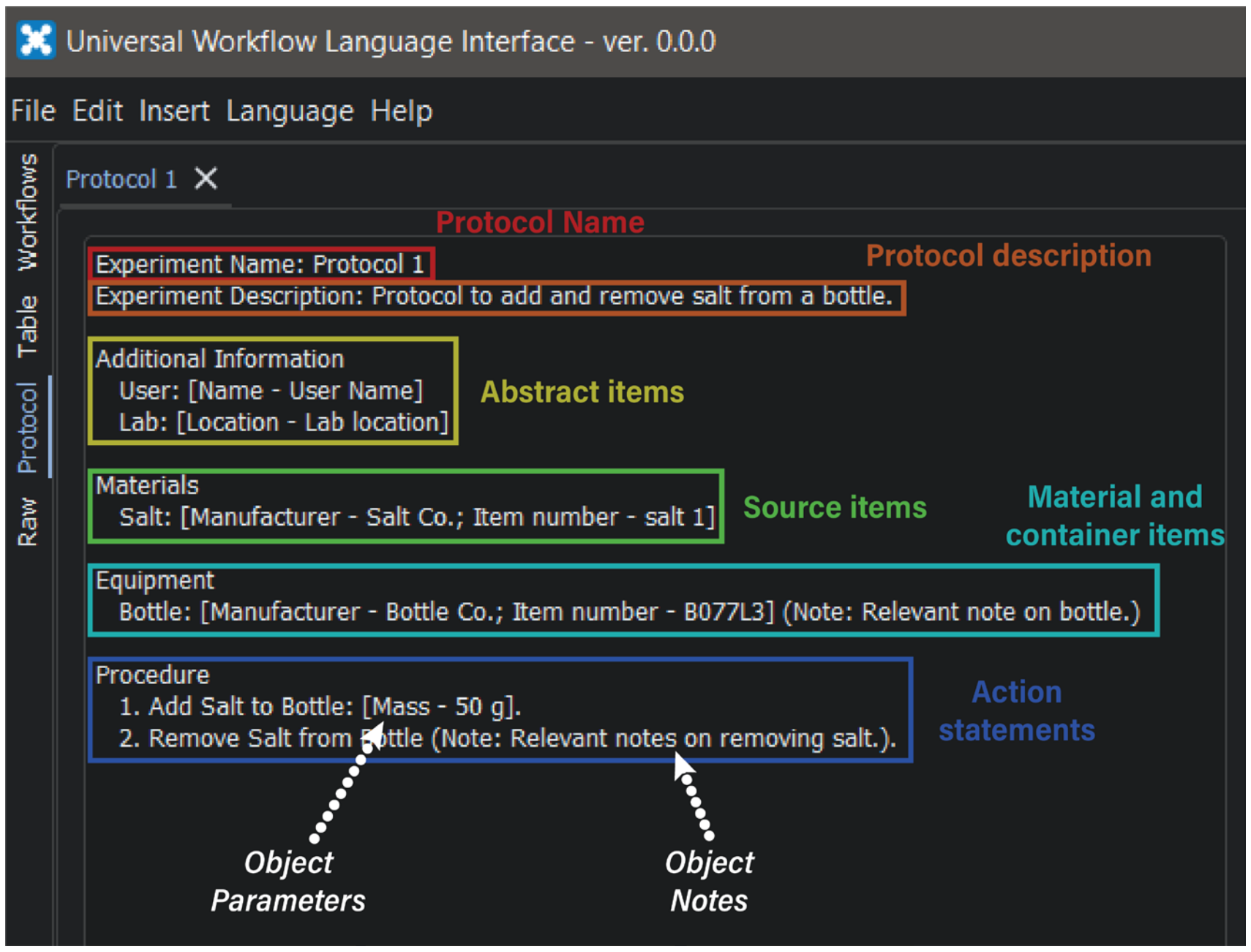

**Figure S.3. Procedurally generated protocol window in the Universal Workflow Language interface with annotations describing the main procedure sections.**

The *Table* tab is capable of both reading and writing to the workflow data. Each parameter within every action and item block is displayed as its own table row with the sections, block name, and parameter name used to identify the row. The value of the parameter is represented by the editable cell in the corresponding row. Changes made in the editable cell are reflected in the workflow. Multiple workflows are represented as additional columns in the table, and overlapping parameter names are displayed on the same row.

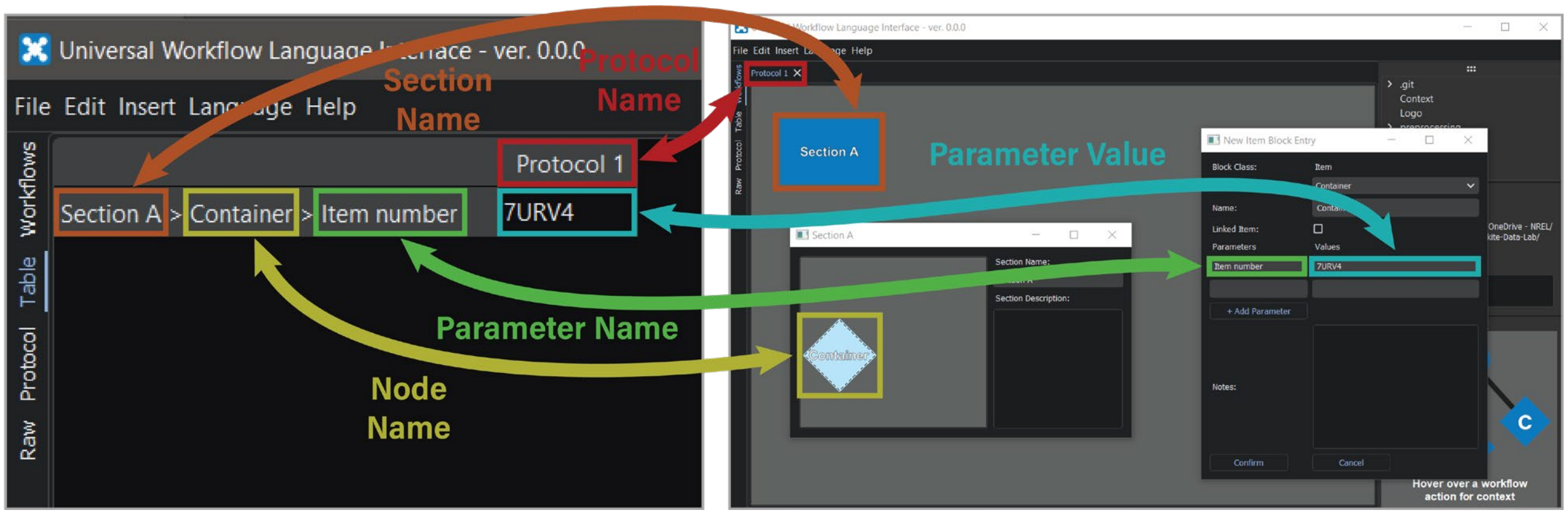

**Figure S.4. Illustration of two-way data flow between the workflow window and the table view window in the Universal Workflow Language interface.**

### *1.10. Interface and Protocol Translation*

Science and research are international endeavors undertaken by a broad and diverse collection of people. Out of necessity, the language of science communication has been constrained primarily to English, but UWL presents a unique opportunity for providing greater flexibility in the languages of protocol reporting and therefore greater precision across diverse inputs. UWL and UWLi are configured for multilingual support across all features of the interface including UWL generation, workspace interaction, data tabulation, and plain text transcriptions. To enable more precise procedure translation, the multilingual support architecture utilizes translation tables over translation models. While language models are constantly improving in accuracy and usability, there remains a considerable limitation in the degree of precision across translations. These imperfections are acceptable for most applications of model-driven translation; however, small errors in a scientific procedure can have significant implications. Translation tables on the other hand allow humans to specify and agree upon direct translations between terms, which allows for clear and reversable translations across languages. To implement translations tables for a topic as large as all scientific procedures, there must be considerable simplification of the parameter space size. UWL achieves this simplification through the inherent breakdown of sentence structures into actions and items. With UWL files, only the action, item, and parameter names are translated from the source language to English before saving.

Compare the UWL approach to the conventional translation of natural language protocols. Implementing a root language in UWL with translation tables simplifies the challenge of translation by reducing the number of routes that must be validated. Instead of attempting to directly translate phrases between every combination of separate languages, all languages interact with a translation table through term and phrase keys. As a result, the total number of translations that must be validated by human researchers scales linearly with the number of languages instead of factorially. Additionally, the isolation of action, item, and parameter terms coupled with the inbuilt logic and structure of UWL protocol steps allows bypassing of full sentence translations. This deconstruction of sentence structure allows for the total translations to scale linearly with the total number of actions, items, and parameters instead of combinatorically.

It should be noted that this work provides the framework for direct translation of protocol elements, and it does not present a fully validated protocol translation system. All translations provided in the current software are generated programmatically using the *Azure Translation API*, and the current iteration has a high probability of providing inaccurate translations in the generated protocol due to the procedural methods used to deconstruct and designate sentence structure. Many written languages contain complexities, such as inseparable verbs or gendered language, that extend beyond the scope of this work and software capabilities. It should also be noted that many of the discussed parameter space simplification methods are inbuilt in existing advanced translation models and algorithms. This discussion is intended to focus on minimizing the labor requirements of providing human validated translations and not on generating higher performing language models.

## 2. Literature Protocol Sampling

**Table S.1. Summary data from literature transcription and analysis.**

| | *Case Study A* | *Case Study B* | *Case Study C* | *Mean* |
|---|---|---|---|---|
| *Words Original* | 267 | 397 | 219 | 294.33 |
| *Materials Included* | 10 | 9 | 3 | 7.33 |
| *Materials Missing* | 2 | 0 | 1 | 1.00 |
| *Equipment Included* | 1 | 7 | 7 | 5.00 |
| *Equipment Missing* | 22 | 32 | 6 | 20.00 |
| *Item Params Included* | 1 | 16 | 6 | 7.67 |
| *Item Params Missing* | 76 | 88 | 40 | 68.00 |
| *Action Params Included* | 15 | 28 | 10 | 17.67 |
| *Action Params Missing* | 28 | 26 | 11 | 21.67 |
| *Ambiguities* | 58 | 63 | 30 | 50.33 |

### *2.1. Literature Case Study A*

**Symmetry breaking and chiral amplification in prebiotic ligation reactions**

Deng, M., Yu, J. and Blackmond, *Nature* **626**, 1019–1024 (2024)

#### *2.1.1. Protocol Excerpts*

The evaluated protocol excerpts were extracted from the supporting information section covering symmetry breaking and chiral amplification in prebiotic ligation reactions and includes additional information available in separate portions of methods. The total work count was calculated from text directly relevant to the specific reaction.

**Legend: Manually Identified Ambiguity**

Page S2:

"…purchased from [1] the following companies…"

Page S3:

"To a [2] solution of $KH_2PO_4$ ([3] 50.0 mmol, 6.80 g) and KCN ([4] 50.0 mmol, 3.16 g) in deionized water ([5] 50 mL), the corresponding aldehyde ([6] 50.0 mmol) [7] was added dropwise at 0 °C. The [8] pH of the reaction was then adjusted to 7 with [9] NaOH solution. The reaction mixture…"

"…[10] until the full consumption of aldehyde."

"…was added [11] dropwise [12] at 0 °C."

"…warm up to [13] room temperature [14] gradually and [15] periodically analyzed…"

"Acetic anhydride ([16] 150 mmol, 14.3 mL 15.3 g) was added [17] dropwise…"

"…at [18] 0 °C…"

"…stirred [19] overnight and allowed to warm up to [20] room temperature [21] gradually."

"…product [22] was extracted…"

"…[23] dried over [24] $MgSO_4$, [25] filtrated, and [26] concentrated."

"…on a [27] silica column…"

### *2.1.2. Protocol Ambiguities*

[1a] Which chemicals were purchased from which company?

[1b] What are the item numbers for the chemicals?

[2a] What were the environmental controls?

[2b] What was the mixing method?

[2c] How long were the materials combined before proceeding to the next step?

[2d] What type of container was the solution prepared in?

[3] What was the measurement method and tool for the $KH_2PO_4$? (i.e. Was the material measured directly into the container, or was it transferred from a weigh boat?)

[4] What was the measurement method and tool for the KCN?

[5] What was the measurement method and tool for the deionized water?

[6] What was the mass conversion basis for the aldehyde? (i.e. What was the actual mass measured from the source container?)

[7a] What was the dripping rate of the solution?

[7b] What equipment was used for dripping?

[7c] Which container was cooled, the aldehyde or the $KH_2PO_4$ / KCN solution?

[7d] What tool was used for cooling?

[7e] What was the cooling rate and time the solution was left at equilibrium?

[8a] What was the rate of addition of the NaOH solution?

[8b] How much NaOH was added total?

[8c] How often was the pH measured? *

* Note: pH equipment and conditions reported elsewhere in manuscript.

[9a] How much NaOH is in the solution?

[9b] How much deionized water is in the solution?

[9c] How were the materials measured?

[9d] What container was the solution prepared in?

[9e] How were the materials mixed?

[9f] How long was the solution left after preparation?

[10a] How often were 1H NMR spectra collected? *

* Note: 1H  NMR equipment and conditions reported elsewhere in manuscript.

[10b] How much of the solution was collected for measurements?

[11a] What was the dripping rate of the solution?

[11b] What equipment was used for dripping?

[12a] How was the mixture cooled?

[12b] How was the mixture mixed?

[12c] How was the temperature measured? (i.e. Was the solution measured from the cooling apparatus or was it measured from a probe in the solution?)

[13a] What is the temperature of the room?

[13b] Was the container left open to ambient?

[14] What was the warming rate? (or) What is the apparatus around the container during warming?

[15a] How often were 1H NMR spectra collected?

[15b] What was the metric to determine when the reaction is stopped?

[16] How was the acetic anhydride measured? (i.e. Was the measurement made based on volume or mass, and what equipment was used?

[17a] What was the dripping rate of the solution?

[17b] What equipment was used for dripping?

[18a] How was the mixture cooled?

[18b] How was the mixture mixed?

[19] How long was the mixture stirred?

[20a] Was the container left open to ambient?

[20b] What is the temperature of the room?

[21a] What was the warming rate?

[21b] Was the solution allowed to warm before or after being left overnight?

[22a] What container was used for the extraction?

[22b] How long and what method was used to mix the materials?

[22c] How were the organic layers removed in the extraction?

[23a] What container was used to dry the solution?

[23b] How long and what method was used to mix the materials?

[24] How much $MgSO_4$ was used to dry the solution?

[25a] What equipment was used for filtration?

[25b] What conditions were used for filtration?

[26a] What equipment was used for concentration?

[26b] What conditions were used for concentration?

[27a] What type of silica and column equipment was used for flash chromatography? *

* Note: Product specific chromatography conditions listed elsewhere in manuscript.

[27b] How was the silica column prepared?

### *2.1.3. UWL Transcribed Procedure*

**Legend: Missing Parameter Missing Item**

Experiment Name: Nature v2
Experiment Description: General Method for Racemic Ac-AA-CN/Ac-AA-13CN Synthesis

Additional Information
Location: [Facility - ###; City - ###; State - ###; Country - ###]
User: [Form Filled by - ###; Experiment conducted by - ###]

Materials
Potassium dihydrogen phosphate: [Supplier - ###; Item number - ###]
Potassium cyanide: [Supplier - ###; Item number - ###]
Corresponding aldehyde: [Supplier - ###; Item number - ###]
Deionized water: [Dispenser manufacturer - ###; Dispenser item number - ###]
Sodium hydroxide: [Supplier - ###; Item number - ###]
Ammonia: [Supplier - ###; Item number - ###; Weight percent - 28%]
Acetic anhydride: [Supplier - ###; Item number - ###]
Ethyl acetate: [Supplier - ###; Item number - ###]
Magnesium sulfate: [Supplier - ###; Item number - ###]
Silica: [Item number - ###; Manufacturer - ###]
Chromatography solvent: [Supplier - ###; Item number - ###]
Nitrogen: [Supplier - ###; Item number - ###]

Equipment
Stir bar: [Manufacturer - ###; Item number - ###]
Scale: [Manufacturer - ###; Item number - ###]
Stir plate: [Manufacturer - ###; Item number - ###]
Weigh boat: [Manufacturer - ###; Item number - ###]
Burette: [Manufacturer - ###; Item number - ###]
Spatula: [Manufacturer - ###; Item number - ###]
Graduated cylinder: [Manufacturer - ###; Item number - ###]
Beaker: [Manufacturer - ###; Item number - ###]
Burette (1): [Manufacturer - ###; Item number - ###]
Cooling apparatus: [Manufacturer - ###; Item number - ###]
Burette (2): [Manufacturer - ###; Item number - ###]
Beaker (1): [Manufacturer - ###; Item number - ###]
Burette (3): [Manufacturer - ###; Item number - ###]
Liquid transfer tool: [Item number - ###; Manufacturer - ###]
Graduated cylinder (1): [Manufacturer - ###; Item number - ###]
Liquid transfer tool (1): [Item number - ###; Manufacturer - ###]
Beaker (2): [Item number - ###; Manufacturer - ###]
Liquid transfer tool (2): [Item number - ###; Manufacturer - ###]

Filtering apparatus: [Item number - ###; Manufacturer - ###]
Concentrating apparatus: [Item number - ###; Manufacturer - ###]
Graduated cylinder (2): [Manufacturer - ###; Item number - ###]
Chromatography column: [Item number - ###; Manufacturer - ###]
Pressurizing apparatus: [Manufacturer - ###; Item number - ###]

Procedure
1. Add Stir bar to Beaker.
2. Add Potassium dihydrogen phosphate to Beaker using Scale, Weigh boat, and Spatula: [Mass - 6.8 g; Moles - 50.0 mmol].
3. Add Potassium cyanide to Beaker using Scale, Weigh boat, and Spatula: [Mass - 3.16 g; Moles - 50.0 mmol].
4. Add Deionized water to Beaker using Graduated cylinder: [Volume - 50 mL].
5. Stir Solution with Stir plate: [Stir rate - ###; Time - ###].
6. Cool Solution with Cooling apparatus: [Temperature - 0 C; Time - ###].
7. Add Corresponding aldehyde to Beaker using Burette: [Drip rate - ###; Volume - ###; Moles - 50.0 mmol].
8. Add Sodium hydroxide to Beaker using Scale and Spatula: [Mass - ###].
9. Add Deionized water to Beaker using Graduated cylinder: [Volume - ###].
10. Pour Sodium hydroxide solution into Burette using Beaker.
11. Add Sodium hydroxide solution to Beaker using Burette: [Drip rate - ###; Volume - ###] (Note: Use the sodium sulfide mixture to adjust the solution to a pH of 7.).
12. Wait : [Time - ###] (Note: Stir until full consumption of the aldehyde. Use 1H NMR to monitor conversion.).
13. Add Ammonia to Beaker using Burette: [Drip rate - ###; Volume - 6.94 mL].
14. Remove Beaker from Cooling apparatus.
15. Wait : [Time - ###] (Note: Allow solution to reach room temperature. Monitor the conversion of the cyanohydrin to the corresponding aminonitrile through 1H NMR.).
16. Cool Solution with Cooling apparatus: [Temperature - 0 C; Time - ###].
17. Add Acetic anhydride to Beaker using Burette: [Moles - 150 mmol; Volume - 14.3 mL; Mass - 15.3 g; Drip rate - ###] (Note: The solution pH was maintained at 9.
### How and when was the pH adjusted?).
18. Remove Beaker from Cooling apparatus.
19. Wait : [Time - ###].
20. Add Ethyl acetate to Beaker using Graduated cylinder: [Volume - 100 mL].
21. Swirl Beaker: [Time - ###; Rate - ###].
22. Transfer Organic phase to Beaker using Liquid transfer tool.
23. Add Ethyl acetate to Beaker using Graduated cylinder: [Volume - 100 mL].
24. Swirl Beaker: [Time - ###; Rate - ###].
25. Transfer Organic phase to Beaker using Liquid transfer tool.
26. Add Ethyl acetate to Beaker using Graduated cylinder: [Volume - 100 mL].
27. Swirl Beaker: [Time - ###; Rate - ###].
28. Transfer Organic phase to Beaker using Liquid transfer tool.

29. Add Magnesium sulfate to Beaker using Scale, Weigh boat, and Spatula: [Mass - ###; Moles - ###].
30. Swirl : [Time - ###; Rate - ###].
31. Transfer Organic phase to Filtering apparatus.
32. Filter Organic phase with Filtering apparatus.
33. Transfer Organic phase to Concentrating apparatus.
34. Concentrate Organic phase with Concentrating apparatus.
35. Add Silica to Chromatography column: [Mass - ###].
36. Transfer Organic phase to Chromatography column.
37. Add Chromatography solvent to Chromatography column using Graduated cylinder: [Volume - ###].
38. Connect Pressurizing apparatus to Chromatography column.
39. Pressurize Chromatography column with Nitrogen using Pressurizing apparatus.
40. Collect Product from Chromatography column.

### *2.2. Literature Case Study B*

**Stable anchoring of single rhodium atoms by indium in zeolite alkane dehydrogenation catalysts**

Lei Zeng, Kang Cheng, Fanfei Sun, Qiyuan Fan, Laiyang Li, Qinghong Zhang, Yao Wei, Wei Zhou, Jincan Kang, Qiuyue Zhang, Mingshu Chen, Qiunan Liu, Liqiang Zhang, Jianyu Huang, Jun Cheng, Zheng Jiang, Gang Fu, and Ye Wang, *Science* **383**, 998–1004 (2024)

#### *2.2.1. Protocol Excerpt*

The protocol excerpt was taken from the supporting information sections on the chemicals and materials and the catalysts synthesis and preparation. The total word count was calculated from all text directly relevant to the catalyst synthesis protocol.

**Legend: Manually Identified Ambiguity**

Supporting information Page 2:

"…prepared by [1] dissolving rhodium (III) chloride hydrate ([2] 1.0 g)…"

"…salts into [3] 10 mL deionized water. Then, [4] 4 mL ethylenediamine…"

"…metal solution [5] under an ice-water bath [6] with stirring [7] until complete dissolution. The final step involved [8] diluting the solution to a total volume of [9] 25 mL."

"…TPAOH solution (25 wt% in water, [10] 5.56 g), deionized water ([11] 5.78 g), and TEOS ([12] 3.47 g) [13] was prepared…"

"…was continuously [14] stirred for 8 hours."

"…carried out in [15] stainless-steel autoclave at [16] 170 °C for 3 days."

"…was recovered by [17] centrifugation, [18] washed with [19] ethanol, and then [20] dried [21] overnight at [22] 80 °C."

"…obtained by [23] calcining the dried powders at [24] 550 °C for 2 hours, with a [25] heating rate of 1.5 °C min–1."

"…TPAOH solution (25 wt%, [26] 2.22 g), TPAOH solution (50 wt%, with 20 % methanol, [27] 1.67 g), TEOS ([28] 3.47 g), and deionized water ([29] 7.78 g)…"

"…[30] an appropriate amount of Rh-thylenediamine complex…"

"…followed by [31] cooling naturally."

"…collected by [32] centrifugation, [33] washed with [34] ethanol, and dried [35] overnight at 60 °C in an [36] oven."

"…obtained by [37] calcining the dried powders at [38] 550 °C for 2 hours…"

"…heating rate of [39] 1.5 °C $min^{-1}$."

"…synthesis stage [40] was controlled at 0.40 wt%…"

### *2.2.2. Protocol Ambiguities*

[1a] How was the solution mixed?

[1b] How long was the solution mixed before dissolving?

[2] What was the measurement method and tool for rhodium (III) chloride hydrate?

[3] What was the measurement method and tool for the deionized water?

[4] What was the measurement method and tool for the ethylenediamine?

[5a] How much water and ice were added to form the bath?

[5b] What container was used for the bath?

[5c] What extent was the reaction vessel submerged?

[5d] How long was the solution allowed to equilibrate?

[6a] How was the solution stirred?

[6b] Under what conditions (e.g. stir rate) was the solution stirred?

[7] How long was it stirred for dissolution?

[8a] How was the solution mixed for dilution?

[8b] Was the solution still under an ice bath?

[9a] How was the dilution volume measured?

[9b] What was the diluting species?

[10] What was the measurement method and tool for the TPAOH?

[11] What was the measurement method and tool for the deionized water?

[12] What was the measurement method and tool for the TEOS?

[13] What were the environmental controls during material loading?

[14a] What method and tool were used to stir the solution?

[14b] What were the stirring conditions such as stir rate during mixing?

[15] Were there any additional controls on the autoclave such as gas flow lines, relief valves, or pressure gauges?

[16a] What tool and conditions were used to heat the autoclave?

[16b] What was the heating rate?

[17a] What equipment was used for centrifugation?

[17b] What settings and time were used for centrifugation?

[17c] What tools and methods were used to extract the solids?

[18] What tools and methods were used to wash the solids?

[19a] What quantity of ethanol was added for washing?

[19b] How was the ethanol measured?

[20] What environmental conditions are used during drying?

[21] How long quantitatively were the solids dried?

[22a] What equipment was used for heating?

[22b] Was the heating tool preheated or is there a relevant ramp rate?

[23a] What environmental conditions were used during calcining?

[23b] What containers were used for calcining the powders?

[24a] What equipment was used for calcining?

[24b] Was the temperature held at 550 $^{O}$C for two hours, or is the ramping time included in the time?

[25] What is the starting temperature before ramping?

[26] What tools were used to weigh the TPAOH?

[27] What tools were used to weigh the methanol?

[28] What tools were used to weigh the TEOS?

[29] What tools were used to weigh the deionized water?

[30] How much solution was added quantitatively for each composition?

[31a] How long was the solution allowed to cool?

[31b] What are the environmental conditions for cooling?

[32a] What equipment and tools were used for centrifugation?

[32b] What conditions were used for centrifugation?

[32c] How was the solid separated remaining liquid solution?

[33a] What method and conditions were used to mix the ethanol with the solution during washing?

[33b] How was the ethanol removed from the solid?

[34] How much ethanol was added for washing?

[35] How long quantitatively were the products dried?

[36a] Was the oven preheated before drying or was a ramp up protocol conducted?

[36b] Were any environmental controls implemented during drying?

[37] What equipment was used for calcining?

[38a] How was the temperature measured?

[38b] Where any environmental controls used during heating?

[38c] When did the timer for two hours start? Was it before or after temperature ramping?

[39a] What was the starting temperature for ramping?

[39b] What tool was used to regulate the ramping?

[40] What equipment and tools were used to measure and add the appropriate quantity of rhodium?

### *2.2.3. UWL Transcribed Procedure*

**Legend: Missing Parameter Missing Item**

Experiment Name: Science v2
Experiment Description:

Additional Information
Location: [Facility - ###; City - ###; State - ###; Country - ###]
User: [Form Filled by - ###; Experiment conducted by - ###]

Materials
Rhodium (III) chloride hydrate: [Supplier - Sinopharm Chemical Reagent Co., Ltd; Item number - ###]
Ice
Water
Ethylenediamine: [Supplier - Sinopharm Chemical Reagent Co., Ltd; Item number - ###]
Deionized water: [Dispenser manufacturer - ###; Dispenser item number - ###]
Tetrapropyl ammonium hydroxide: [Item number - ###; Supplier - Sinopharm Chemical Reagent Co., Ltd; Weight percent - 25%; Balance - Water]
Ethanol: [Supplier - ###; Item number - ###]
Tetraethyl orthosilicate: [Item number - ###; Supplier - Sinopharm Chemical Reagent Co., Ltd]
Tetrapropyl ammonium hydroxide (1): [Item number - ###; Supplier - Aladdin Industrial Inc.; Weight percent - 50%; Balance - Water; Weight percent methanol - 20%]

Equipment
Stir bar: [Manufacturer - ###; Item number - ###]
Weigh boat: [Manufacturer - ###; Item number - ###]
Spatula: [Manufacturer - ###; Item number - ###]
Graduated cylinder: [Manufacturer - ###; Item number - ###]
Ice bath: [Manufacturer - Xiamen University Glassware Workshop; Item number - ###]
Beaker: [Manufacturer - Xiamen University Glassware Workshop; Item number - ###]
Stir plate: [Manufacturer - ###; Item number - ###]
Graduated cylinder (1): [Manufacturer - ###; Item number - ###]
Teflon liner: [Manufacturer - ###; Item number - ###]
Stir bar (1): [Manufacturer - ###; Item number - ###]
Autoclave: [Manufacturer - ###; Item number - ###]
Heating apparatus: [Manufacturer - ###; Item number - ###]
Mixture transfer tool: [Manufacturer - ###; Item number - ###]
Scale: [Manufacturer - ###; Item number - ###]
Liquid removal tool: [Manufacturer - ###; Item number - ###]
Falcon tube: [Manufacturer - ###; Item number - ###]
Spatula (1): [Manufacturer - ###; Item number - ###]
Graduated cylinder (2): [Manufacturer - ###; Item number - ###]
Beaker (1): [Manufacturer - Xiamen University Glassware Workshop; Item number - ###]
Liquid removal tool (1): [Manufacturer - ###; Item number - ###]
Spatula (2): [Manufacturer - ###; Item number - ###]

Oven: [Manufacturer - ###; Item number - ###]
Beaker (2): [Manufacturer - ###; Item number - ###]
Stir bar (2): [Manufacturer - ###; Item number - ###]
Teflon liner (1): [Manufacturer - ###; Item number - ###; Volume - 100 mL]
Graduated cylinder (3): [Manufacturer - ###; Item number - ###]
Magnetic stirrer: [Manufacturer - ###; Item number - ###]
Autoclave (1): [Manufacturer - ###; Item number - ###]
Mixture transfer tool (1): [Manufacturer - ###; Item number - ###]
Centrifuge: [Manufacturer - ###; Item number - ###]
Falcon tube (1): [Manufacturer - ###; Item number - ###]
Oven (1): [Manufacturer - ###; Item number - ###]
Liquid removal tool (2): [Manufacturer - ###; Item number - ###]
Spatula (3): [Manufacturer - ###; Item number - ###]
Graduated cylinder (4): [Manufacturer - ###; Item number - ###]
Beaker (3): [Manufacturer - Xiamen University Glassware Workshop; Item number - ###]
Liquid removal tool (3): [Manufacturer - ###; Item number - ###]
Spatula (4): [Manufacturer - ###; Item number - ###]
Beaker (4): [Manufacturer - Xiamen University Glassware Workshop; Item number - ###]

Procedure
1. Add Stir bar to Beaker.
2. Add Deionized water to Beaker using Graduated cylinder: [Volume - 10 mL].
3. Add Rhodium (III) chloride hydrate to Beaker using Weigh boat, Spatula, and Scale: [Mass - 1.0 g].
4. Add Ice to Ice bath: [Mass - ###].
5. Add Water to Ice bath: [Volume - ###].
6. Transfer Beaker to Ice bath.
7. Stir Solution with Stir plate: [Stir rate - ###].
8. Add Ethylenediamine to Beaker using Graduated cylinder: [Volume - 10 mL].
9. Wait : [Time - ###].
10. Add Deionized water to Beaker using Graduated cylinder: [Volume - 11 mL].
11. Remove Beaker from Ice bath.
12. Add Teflon liner to Autoclave.
13. Add Stir bar to Autoclave.
14. Add Tetrapropyl ammonium hydroxide to Autoclave using Scale: [Mass - 5.56 g].
15. Add Deionized water to Autoclave using Scale: [Mass - 5.78 g].
16. Add Tetraethyl orthosilicate to Autoclave using Scale: [Mass - 3.47 g].
17. Stir Autoclave: [Stir Rate - ###; Time - 8 hr].
18. Heat Autoclave with Heating apparatus: [Temperature - 170 C; Ramp rate - ###].
19. Wait : [Time - 3 days].
20. Remove Mixture from Autoclave using Mixture transfer tool.
21. Transfer Mixture to Falcon tube using Mixture transfer tool: [Volume - ###; Parts - ###].
22. Centrifuge Falcon tube: [Rate - ###; Time - ###].
23. Remove Supernatant from Falcon tube using Liquid removal tool.
24. Remove Precipitate from Falcon tube using Spatula.
25. Add Precipitate to Beaker.

26. Add Ethanol to Beaker using Graduated cylinder: [Volume - ###].
27. Swirl Beaker: [Time - ###; Rate - ###].
28. Remove Liquid from Beaker using Liquid removal tool.
29. Remove Solid from Beaker using Spatula.
30. Transfer Solid to Beaker using Spatula.
31. Transfer Beaker to Oven.
32. Heat Beaker using Oven: [Temperature - 80 C].
33. Wait : [Time - ###].
34. Heat Beaker using Oven: [Temperature - 550 C; Ramp rate - 1.5 C/min].
35. Wait : [Time - 2 hr].
36. Remove Beaker from Oven.
37. Cool Beaker: [Temperature - Ambient; Time - ###].
38. Add Stir bar to Teflon liner.
39. Add Tetrapropyl ammonium hydroxide to Teflon liner using Scale: [Mass - 2.22 g].
40. Add Tetrapropyl ammonium hydroxide to Teflon liner using Scale: [Mass - 1.67 g].
41. Add Tetraethyl orthosilicate to Teflon liner: [Mass - 3.47 g].
42. Add Deionized water to Teflon liner: [Mass - 7.78 g].
43. Add Rh-thylenediamine complex to Teflon liner using Graduated cylinder: [Volume - ###] (Note: The content of rhodium in the synthesis stage was controlled at 0.40 wt% ).
44. Stir Teflon liner with Magnetic stirrer: [Rate - 400 rpm; Time - 8 hr].
45. Transfer Teflon liner to Autoclave.
46. Transfer Autoclave to Oven.
47. Heat Autoclave using Oven: [Temperature - 170 C; Time - 3 days].
48. Remove Autoclave from Oven.
49. Cool Autoclave: [Temperature - Ambient; Time - ###].
50. Remove Mixture from Autoclave using Mixture transfer tool.
51. Transfer Mixture to Falcon tube using Mixture transfer tool: [Volume - ###; Parts - ###].
52. Transfer Falcon tube to Centrifuge.
53. Centrifuge Falcon tube: [Rate - ###; Time - ###].
54. Remove Falcon tube from Centrifuge.
55. Remove Supernatant from Falcon tube using Liquid removal tool.
56. Remove Precipitate from Falcon tube using Spatula.
57. Add Precipitate to Beaker.
58. Add Ethanol to Beaker using Graduated cylinder: [Volume - ###].
59. Swirl Beaker: [Time - ###; Rate - ###].
60. Remove Liquid from Beaker using Liquid removal tool.
61. Remove Solid from Beaker using Spatula.
62. Transfer Solid to Beaker using Spatula.
63. Transfer Beaker to Oven.
64. Heat Beaker using Oven: [Temperature - 60 C].
65. Wait : [Time - ###].
66. Heat Beaker using Oven: [Temperature - 550 C; Ramp rate - 1.5 C/min].
67. Wait : [Time - 2 hr].
68. Remove Beaker from Oven.
69. Cool Beaker: [Temperature - Ambient; Time - ###].

### *2.3. Literature Case Study C*

**Smart-Responsive HOF Heterostructures with Multiple Spatial-Resolved Emission Modes toward Photonic Security Platform**

Yuanchao Lv, Jiashuai Liang, Zhile Xiong, Xue Yang, Yunbin Li, Hao Zhang, Shengchang Xiang, Banglin Chen, and Zhangjing Zhang, *Adv. Mater.* 36, 230913 (2024)

#### *2.3.1. Protocol Excerpt*

The protocol excerpts were taken from the supporting information on sections covering the materials, HOF-FHU-39 synthesis, and HOF heterostructure synthesis. The total word count was calculated from the text in the supporting information relevant to these two syntheses.

**Legend: Manually Identified Ambiguity**

Supporting Information Page 4:

"…solvents were [1] dried and [2] distilled before use."

"…the o-tfpe ([3] 0.005 g, 0.0069 mmol) was dissolved in DMF ([4] 5 mL)…"

"…to slowly evaporate in a [5] loosely capped vial over 12 hours, at [6] atmospheric pressure and [7] room temperature…"

"…m-tfpe ([8] 2.5 mg) was dissolved in DMF ([9] 5 mL)…"

"Then, [10] 400 µL of the m-tfpe solution was [11] dropcast…"

"…for HOF-FJU-40 [12] was carried out in a weighing bottle…"

"…glass cap at [13] room temperature."

"Subsequently, the [14] 400 µL of o-tfpe solution in DMF ([15] 1 mg/ml) was added…"

"…with cyclohexane [16] as a counter-solvent atmosphere to trigger…"

"…were obtained after [17] slow evaporation of the solvent…"

### *2.3.2. Protocol Ambiguities*

[1a] What materials were used for drying?

[1b] What equipment was used for drying?

[1c] What methods and procedures were used for drying?

[2a] What equipment was used for distilling?

[2b] What methods and procedures were used for distilling?

[3] What was the measurement method and tool used for o-tfpe?

[4] What was the measurement method and tool used for DMF?

[5] How specifically or quantitatively was the vial loosely capped? (e.g. What is the size of the gap left open to ambient air after capping?)

[6a] What are the environmental conditions around the vial constituting atmospheric pressure? (e.g. Was the vial in a region with air flow, such as in a fume hood? Were any environmental controls implemented or was the solution open to ambient air?)

[6b] If listing atmospheric pressure, where and was the experiment conducted?

[7a] What was the temperature of the room?

[7b] How was the temperature measured?

[8] What was the measurement method and tool used for m-tfpe?

[9] What was the measurement method and tool used for DMF?

[10] What was the measurement method and tool used for the m-tfpe solution?

[11a] What was the injection rate for drop casting?

[11b] What was the position of the injection tool during drop casting?

[12a] Was the glass substrate transferred to the weighing bottle before or after drop casting?

[12b] How was the substrate held above the count-solvent level in the weighing bottle?

[13] What was the temperature of the room?

[14a] What was the measurement method and tool used for the o-tfpe solution?

[14b] What was the injection rate for drop casting?

[14c] What was the position of the injection tool during drop casting?

[15a] How was this solution prepared? (e.g. Was this mixed for a specific duration of time? What were the conditions during combination?)

[15b] What tools were used to measure the quantity of o-tfpe?

[15c] What tools were used to measure the quantity of DMF?

[16a] Was the cyclohexane added before or after drop casting?

[16b] How much cyclohexane was added to the container?

[16c] What materials and methods were used to add cyclohexane to the container?

[17] How and with what parameters was the evaporation rate regulated?

### *2.3.3. UWL Transcribed Procedure*

**Legend: Missing Parameter Missing Item**

Experiment Name: Advanced Materials v2
Experiment Description:

Additional Information
Location: [Facility - ###; City - ###; State - ###; Country - ###]
Room: [Temperature - ###; Pressure - ###; Humidity - ###]
User: [Form Filled by - ###; Synthesis conducted by - ###]

Materials
1,1,2,2-tetrakis(4-(6-fluoropyridin-3-yl)phenyl)ethene (Note: Synthesized through classical Suzuki condensation reaction.)
Dimethylformamide: [Supplier - ###; Item number - ###]
1,1,2,2-tetrakis(4-(5- fluoropyridin-3-yl)phenyl)ethene (Note: Synthesized through classical Suzuki condensation reaction.)
Cyclohexane: [Supplier - ###; Item number - ###]

Equipment
Graduated cylinder: [Manufacturer - ###; Item number - ###]
Scale: [Manufacturer - ###; Item number - ###]
Vial: [Manufacturer - ###; Item number - ###; Volume - 10 mL]
Spatula: [Manufacturer - ###; Item number - ###]
Cap: [Manufacturer - ###; Item number - ###]
Graduated cylinder (1): [Manufacturer - ###; Item number - ###]
Vial (1): [Manufacturer - ###; Item number - ###; Volume - 10 mL]
Cap (1): [Manufacturer - ###; Item number - ###]
Pipettor: [Manufacturer - ###; Item number - ###]
Substrate: [Manufacturer - ###; Item number - ###; Length - 22 mm; Width - 22 mm; Height - ###]
Weigh bottle: [Manufacturer - ###; Item number - ###; Length - 30 mm; Diameter - 70 mm]
Graduated cylinder (2): [Manufacturer - ###; Item number - ###]
Glass cap: [Manufacturer - ###; Item number - ###]

Procedure
1. Add 1,1,2,2-tetrakis(4-(6-fluoropyridin-3-yl)phenyl)ethene to Vial using Scale and Spatula: [Mass - 0.005 g; Moles - 0.0069 mmol].
2. Add Dimethylformamide to Vial using Graduated cylinder: [Volume - 5 mL].
3. Swirl Vial: [Rate - ###; Time - ###].
4. Cover Vial with Cap (Note: Loosely cover vial with cap).

5. Wait : [Time - 12 hr].
6. Add 1,1,2,2-tetrakis(4-(5- fluoropyridin-3-yl)phenyl)ethene to Vial using Scale and Spatula: [Mass - 2.5 mg].
7. Add Dimethylformamide to Vial using Graduated cylinder: [Volume - 5 mL].
8. Swirl Vial: [Rate - ###; Time - ###].
9. Cover Vial with Cap (Note: Loosely cover vial with cap).
10. Wait : [Time - 12 hr].
11. Add Substrate to Weigh bottle.
12. Drop cast m-tfpe solution onto Substrate using Pippettor: [Volume - 400 uL; Deposition rate - ###; Deposition location - ###].
13. Add Glass cap to Weigh bottle.
14. Wait : [Time - ###].
15. Add Cyclohexane to Weigh bottle using Graduated cylinder: [Volume - ###; Location - ###].
16. Drop cast o-tfpe solution onto Substrate using Pippettor: [Volume - 400 uL; Deposition rate - ###; Deposition location - ###].
17. Add Glass cap to Weigh bottle.
18. Wait : [Time - 12 hr].

### 3. Surrogate Modeling

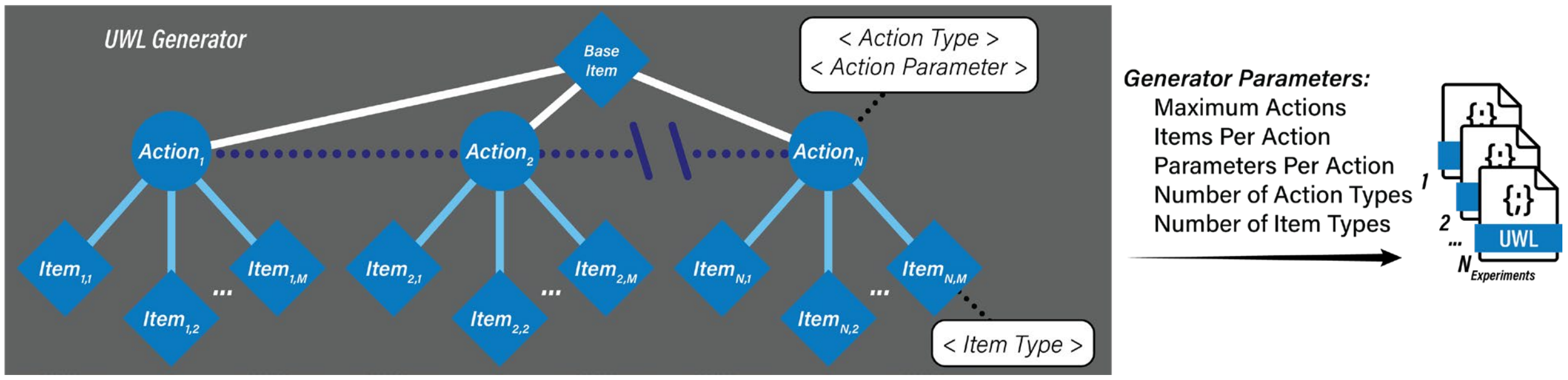

**Figure S.5. Illustration of sample generator for surrogate modeling study. The workflow generator creates a randomly generated data set of UWL files using a specified set of parameter constraints. The workflows are a selection of action steps with a length within a specified number of maximum actions that each contain a continuous numerical parameter and are connected to a specific number of items. The number of possible item types and possible action types are also specified as generator parameters.**

**Table S.1. UWL file generator parameter sets.**

| | Parameters | | | | |
|---|---|---|---|---|---|
| Set | *Max Actions* | *Items per Action* | *Scalar Params per Action* | *Action Types* | *Item Types* |
| *Low* | 3 | 1 | 1 | 5 | 3 |
| *Mid* | 5 | 3 | 1 | 7 | 5 |
| *High* | 7 | 5 | 1 | 9 | 7 |

**Table S.2. UWL Generator metrics for three tested complexity levels.**

| | *Metrics* | | | |
|---|---|---|---|---|
| *Complexity Level* | Max Categorical Params | Max Scalar Params | Max Total Params | Total Categorical Combinations |
| Low | 6 | 3 | 9 | $10^{3}$ |
| Mid | 20 | 5 | 25 | $10^{12}$ |
| High | 42 | 7 | 49 | $10^{25}$ |

**Algorithm S.1. Surrogate function for simulating the effects of multistep experimental procedures for modeling studies.**

***Input***: Action Sequence = $\begin{Bmatrix} 0 \le AP_{i,j} \text{ [Action parameter } j \text{ at step } i] \le 1 \\ AN_i \text{ [Name of action step } i] \in \{1, 2 \dots N_{PAN}\} \\ IN_{i,k} \text{ [Name of item } k \text{ for action step } i] \in \{1, 2 \dots N_{PIN}\} \end{Bmatrix}$

***Output***: $r_{NA}$ [Final response]

***Variables***:
$N_A$ [Number of total action steps]
$N_{AP}$ [Number of action parameters per action]
$N_{PAN}$ [Number of possible action names]
$N_{IN}$ [Number of item names per action]
$N_{PIN}$ [Number of possible item names]
$APF_{i,j}$ [Action parameter feature $j$ at step $i$]
$ASB_i$ [Action name sequence bonus feature at step $i$]
$ANF_i$ [Action name feature at step $i$]
$INF_{i,k}$ [Item name feature k at step $i$]

---

$r_0 = 1$
**for** $i = 1$ *to* $N_A$ **do**
    $APF_{i,0} = 1$
    **for** $j = 1$ *to* $N_{AP}$ **do**
        $APF_{i,j} = APF_{i,j-1}(1 - |0.75 - AP_{i,j}|)$
    **if** $AN_i \in \{1,2,3\}$ **do**
        **if** $(AN_{i-1} = 1)$ *and* $(AN_{i-2} = 2)$ *and* $(AN_{i-3} = 3)$ **do**
            $ASB_i = 0.15$
        **else if** $(AN_{i-1} = 1)$ *and* $(AN_{i-2} = 2)$ **do**
            $ASB_i = 0.10$
        **else if** $(AN_{i-1} = 1)$ **do**
            $ASB_i = 0.05$
        **else**
            $ASB_i = 0$
        $ANF_i = \frac{AN_i}{10}$
        $INF_{0,i} = 1$
        **for** $k = 1$ *to* $N_{IN}$ **do**
            $INF_{k,i} = INF_{k-1,i}\frac{IN_{i,k}}{10}$
        $r_i = ASB_i + \left(APF_{NAP,i}(ANF_i + INF_{NIN,i})\right)^{r_{i-1}}$
    **else if** $AN_i \notin \{1,2,3\}$ **do**
        $r_i = r_{i-1} + \frac{APF_{NAP,i}}{50}$